\RequirePackage[left,columnwise]{lineno}
\documentclass[aps,prc,twocolumn,superscriptaddress]{revtex4-1}

\usepackage{lipsum}
\usepackage{graphicx}
\usepackage{verbatim}
\usepackage{enumitem}
\usepackage{subfigure}
\usepackage{amsmath,amssymb,amsfonts,eufrak,mathrsfs}
\usepackage{bm}
\usepackage{color}
\usepackage[bookmarks,
                bookmarksopen = true,
                bookmarksnumbered = true,
                linktocpage,
                colorlinks = true,
                linkcolor = blue,
                urlcolor  = blue,
                citecolor = blue,
                anchorcolor = green,
                hyperindex = true,
                hyperfigures]
                {hyperref}

\usepackage{lineno}
\usepackage{hyperref}

\graphicspath{{./}}

\newcommand{\CommentBlock}[1]{}

\definecolor{darkgreen}{RGB}{50,150,50}

\def\Ybf{\mathbf{Y}}
\def\Vbf{\mathbf{V}}
\def\Sbf{\mathbf{S}}

\def\thetaE{\theta}
\def\N{\mathsf{N}}

\newcommand{\Qzero}{\ensuremath{Q_0}}

\newcommand{\Matter}{\textsc{Matter}}
\newcommand{\Lbt}{\textsc{Lbt}}

\newcommand{\AuAu}{\ensuremath{\mathrm{Au+Au}}}
\newcommand{\PbPb}{\ensuremath{\mathrm{Pb+Pb}}}

\newcommand{\pp}{\ensuremath{\mathrm{pp}}}

\newcommand{\alphas}{\ensuremath{\alpha_\mathrm{s}}}

\newcommand{\aaa}{\ensuremath{A+A}}
\newcommand{\sqrtsNN}{\ensuremath{\sqrt{s_{\mathrm {NN}}}}}

\newcommand{\gev}{\ensuremath{\mathrm{GeV/}c}}

\newcommand{\TAAavg}{\ensuremath{\langle{T_{AA}}\rangle}}

\newcommand{\pT}{\ensuremath{p_\mathrm{T}}}

\newcommand{\pTh}{\ensuremath{p_{\mathrm{T,h}}}}
\newcommand{\pTjet}{\ensuremath{p_{\mathrm{T,jet}}}}

\newcommand{\qhat}{\ensuremath{\hat{q}}}
\newcommand{\qhatg}{\ensuremath{\hat{q}_g}}

\newcommand{\RAA}{\ensuremath{R_\mathrm{AA}}}

\begin{document}

\newcommand{\red}[1]{{\color{red} #1}}


\title{Determining the jet transport coefficient \qhat\ from inclusive hadron suppression measurements using Bayesian parameter estimation}


\author{S.~Cao}
\affiliation{Department of Physics and Astronomy, Wayne State University, Detroit MI 48201.}
\affiliation{Institute of Frontier and Interdisciplinary Science, Shandong University, Qingdao, Shandong 266237, China}

\author{Y.~Chen}
\affiliation{Laboratory for Nuclear Science, Massachusetts Institute of Technology, Cambridge MA 02139.}

\author{J.~Coleman}
\affiliation{Department of Statistical Science, Duke University, Durham NC 27708.}

\author{J.~Mulligan}
\affiliation{Department of Physics, University of California, Berkeley CA 94270.}
\affiliation{Nuclear Science Division, Lawrence Berkeley National Laboratory, Berkeley CA 94270.}

\author{P.~M.~Jacobs}
\affiliation{Department of Physics, University of California, Berkeley CA 94270.}
\affiliation{Nuclear Science Division, Lawrence Berkeley National Laboratory, Berkeley CA 94270.}

\author{R.~A.~Soltz}
\affiliation{Department of Physics and Astronomy, Wayne State University, Detroit MI 48201.}
\affiliation{Lawrence Livermore National Laboratory, Livermore CA 94550.}

\author{A.~Angerami}
\affiliation{Lawrence Livermore National Laboratory, Livermore CA 94550.}

\author{R. Arora}
\affiliation{University of Texas at San Antonio, 1 UTSA Circle, San Antonio, TX 78249.}

\author{S.~A.~Bass}
\affiliation{Department of Physics, Duke University, Durham NC 27708.}

\author{L.~Cunqueiro}
\affiliation{Department of Physics and Astronomy, University of Tennessee, Knoxville TN 37996.}
\affiliation{Physics Division, Oak Ridge National Laboratory, Oak Ridge TN 37830.}

\author{T.~Dai}
\affiliation{Department of Physics, Duke University, Durham NC 27708.}

\author{L.~Du}
\affiliation{Department of Physics, The Ohio State University, Columbus OH 43210.}

\author{R.~Ehlers}
\affiliation{Department of Physics and Astronomy, University of Tennessee, Knoxville TN 37996.}
\affiliation{Physics Division, Oak Ridge National Laboratory, Oak Ridge TN 37830.}

\author{H.~Elfner}
\affiliation{GSI Helmholtzzentrum f\"{u}r Schwerionenforschung, 64291 Darmstadt, Germany.}
\affiliation{Institute for Theoretical Physics, Goethe University, 60438 Frankfurt am Main, Germany.}
\affiliation{Frankfurt Institute for Advanced Studies, 60438 Frankfurt am Main, Germany.}

\author{D.~Everett}
\affiliation{Department of Physics, The Ohio State University, Columbus OH 43210.}

\author{W.~Fan}
\affiliation{Department of Physics, Duke University, Durham NC 27708.}

\author{R.~J.~Fries}
\affiliation{Cyclotron Institute, Texas A\&M University, College Station TX 77843.}
\affiliation{Department of Physics and Astronomy, Texas A\&M University, College Station TX 77843.}

\author{C.~Gale}
\affiliation{Department of Physics, McGill University, Montr\'{e}al QC H3A-2T8.}

\author{F.~Garza}
\affiliation{Cyclotron Institute, Texas A\&M University, College Station TX 77843.}
\affiliation{Department of Physics and Astronomy, Texas A\&M University, College Station TX 77843.}

\author{Y.~He}
\affiliation{Key Laboratory of Quark and Lepton Physics (MOE) and Institute of Particle Physics, Central China Normal University, Wuhan 430079, China.}

\author{M.~Heffernan}
\affiliation{Department of Physics, McGill University, Montr\'{e}al QC H3A-2T8.}

\author{U.~Heinz}
\affiliation{Department of Physics, The Ohio State University, Columbus OH 43210.}

\author{B.~V.~Jacak}
\affiliation{Department of Physics, University of California, Berkeley CA 94270.}
\affiliation{Nuclear Science Division, Lawrence Berkeley National Laboratory, Berkeley CA 94270.}

\author{S.~Jeon}
\affiliation{Department of Physics, McGill University, Montr\'{e}al QC H3A-2T8.}

\author{W.~Ke}
\affiliation{Department of Physics, University of California, Berkeley CA 94270.}
\affiliation{Nuclear Science Division, Lawrence Berkeley National Laboratory, Berkeley CA 94270.}

\author{B.~Kim}
\affiliation{Cyclotron Institute, Texas A\&M University, College Station TX 77843.}
\affiliation{Department of Physics and Astronomy, Texas A\&M University, College Station TX 77843.}

\author{M.~Kordell~II}
\affiliation{Cyclotron Institute, Texas A\&M University, College Station TX 77843.}

\author{A.~Kumar}
\affiliation{Department of Physics and Astronomy, Wayne State University, Detroit MI 48201.}

\author{A.~Majumder}
\affiliation{Department of Physics and Astronomy, Wayne State University, Detroit MI 48201.}

\author{S.~Mak}
\affiliation{Department of Statistical Science, Duke University, Durham NC 27708.}

\author{M.~McNelis}
\affiliation{Department of Physics, The Ohio State University, Columbus OH 43210.}

\author{C.~Nattrass}
\affiliation{Department of Physics and Astronomy, University of Tennessee, Knoxville TN 37996.}

\author{D.~Oliinychenko}
\affiliation{Nuclear Science Division, Lawrence Berkeley National Laboratory, Berkeley CA 94270.}

\author{C. Park}
\affiliation{Department of Physics, McGill University, Montr\'{e}al QC H3A-2T8.}
\affiliation{Department of Physics and Astronomy, Wayne State University, Detroit MI 48201.}

\author{J.-F. Paquet}
\affiliation{Department of Physics, Duke University, Durham NC 27708.}

\author{J.~H.~Putschke}
\affiliation{Department of Physics and Astronomy, Wayne State University, Detroit MI 48201.}

\author{G.~Roland}
\affiliation{Department of Physics, Massachusetts Institute of Technology, Cambridge MA 02139.}

\author{A.~Silva}
\affiliation{Department of Physics and Astronomy, University of Tennessee, Knoxville TN 37996.}

\author{B.~Schenke}
\affiliation{Department of Physics, Brookhaven National Laboratory, Upton NY 11973.}

\author{L.~Schwiebert}
\affiliation{Department of Computer Science, Wayne State University, Detroit MI 48202.}

\author{C.~Shen}
\affiliation{Department of Physics and Astronomy, Wayne State University, Detroit MI 48201.}
\affiliation{RIKEN BNL Research Center, Brookhaven National Laboratory, Upton NY 11973.}

\author{C.~Sirimanna}
\affiliation{Department of Physics and Astronomy, Wayne State University, Detroit MI 48201.}

\author{Y.~Tachibana}
\affiliation{Department of Physics and Astronomy, Wayne State University, Detroit MI 48201.}
\affiliation{Akita International University, Yuwa, Akita-city 010-1292, Japan}

\author{G.~Vujanovic}
\affiliation{Department of Physics and Astronomy, Wayne State University, Detroit MI 48201.}

\author{X.-N.~Wang}
\affiliation{Key Laboratory of Quark and Lepton Physics (MOE) and Institute of Particle Physics, Central China Normal University, Wuhan 430079, China.}
\affiliation{Department of Physics, University of California, Berkeley CA 94270.}
\affiliation{Nuclear Science Division, Lawrence Berkeley National Laboratory, Berkeley CA 94270.}

\author{R.~L.~Wolpert}
\affiliation{Department of Statistical Science, Duke University, Durham NC 27708.}

\author{Y.~Xu}
\affiliation{Department of Physics, Duke University, Durham NC 27708.}


\collaboration{The JETSCAPE Collaboration}

\date{\today}


\begin{abstract}

We report a new determination of \qhat, the jet transport coefficient of the Quark-Gluon Plasma. We use the JETSCAPE framework, which incorporates a novel multi-stage theoretical approach to in-medium jet evolution and Bayesian inference for parameter extraction. The calculations, based on the \Matter\ and \Lbt\ jet quenching models, are compared to experimental measurements of inclusive hadron suppression in \AuAu\ collisions at RHIC and \PbPb\ collisions at the LHC. The correlation of experimental systematic uncertainties is accounted for in the parameter extraction. The functional dependence of \qhat\ on jet energy or virtuality and medium temperature is based on a perturbative picture of in-medium scattering, with components reflecting the different regimes of applicability of \Matter\ and \Lbt. In the multi-stage approach, the switch between \Matter\ and \Lbt\ is governed by a virtuality scale \Qzero. Comparison of the posterior model predictions to the RHIC and LHC hadron suppression data shows reasonable agreement, with moderate tension in limited regions of phase space. The distribution of $\qhat/T^3$ extracted from the posterior distributions exhibits weak dependence on jet momentum and medium temperature $T$, with 90\% Credible Region (CR) depending on the specific choice of model configuration. The choice of \Matter+\Lbt, with switching at virtuality \Qzero, has 90\% CR of $2<\qhat/T^3<4$ for $\pTjet>40$ \gev. The value of \Qzero, determined here for the first time, is in the range 2.0-2.7 GeV.

\end{abstract}

\maketitle


\section{Introduction}
\label{sec:introduction}

The Quark-Gluon Plasma (QGP) is the state of matter in conditions of extreme temperature and density, similar to those of the universe a few microseconds after the Big Bang, with structure and dynamics governed by interactions of sub-hadronic quanta~\cite{Shuryak:2014zxa}. The QGP is generated in the laboratory by collisions of heavy nuclei (\aaa) at the Relativistic Heavy Ion Collider (RHIC) and the Large Hadron Collider (LHC), and its properties have been measured extensively by large experiments at those facilities (\cite{Busza:2018rrf} and references therein). Comparison of these measurements with theoretical calculations show that the QGP exhibits collective behavior with very low specific viscosity \cite{Heinz:2013th}. The QGP is likewise found to be opaque to penetrating probes carrying color charge, a phenomenon known as ``jet quenching"~\cite{Qin:2015srf}. 

Jets in high-energy hadronic collisions are generated by hard (high $Q^2$) interactions of partons (quarks and gluons) from the incoming projectiles. The scattered partons are initially highly virtual and evolve through QCD bremsstahlung and pair production, forming a parton shower that hadronizes into a collimated spray of stable particles that is observable experimentally (a ``jet''). Jet production rates and jet structure have been measured extensively in $p(\bar{p})+p$ collisions~\cite{ppUA1, ppUA2, ppCDF, ppD0, ppSTAR, ppATLAS7TeV, PhysRevC.101.034911, Sirunyan:2020uoj}, and theoretical calculations based on perturbative QCD (pQCD) are in excellent agreement with such measurements ~\cite{Dasgupta2016, Pires2017, Ringer2018, Czakon2019, Larkoski_2020}.

Jets are likewise produced in high-energy \aaa\ collisions, in parallel with formation of the QGP. Jets generated in \aaa\ collisions propagate during their shower evolution through the QGP and interact with it, thereby modifying final-state jet structure and jet distributions relative to production in vacuum. Measurements of these modifications provide unique and sensitive probes of the QGP.

A key experimental observable of jet quenching is the suppression of inclusive hadron production at high transverse momentum (high \pT)~\cite{Wang:1991xy, Qin:2015srf}. Hadron suppression is measured using the ratio \RAA\ of the inclusive hadron yield  in \aaa\ collisions to that in a reference system, usually \pp\ collisions at the same collision energy, whose yield is scaled to account for nuclear geometry~\cite{Miller:2007ri}; $\RAA=1$ corresponds to the absence of nuclear effects in high-\pT\ hadron production. Such effects arise from both nuclear Parton Distributions Functions and from jet quenching in the QGP. Inclusive hadron suppression at high \pT\ has been measured extensively at RHIC~\cite{Adams:2003kv,Adare:2012wg} and the LHC~\cite{CMS:2012aa,Aad:2015wga,Khachatryan:2016odn,Acharya:2018qsh} .  

Jet quenching is understood theoretically to arise from elastic and inelastic interactions of the partons in the jet shower as it traverses the QGP, with coherence effects playing an important role~\cite{Braaten:1991we, Wang:1991xy, Gyulassy:1993hr, Baier:1996kr, Zakharov:1996fv, Qin:2007rn, Qin:2015srf}. Several different formalisms have been developed to describe this process, as reviewed in~\cite{Armesto:2011ht}. Calculations based on these formalisms have been carried out for  inclusive hadron suppression~\cite{Bass:2008rv, Armesto:2009zi, Chen:2010te,Cao:2017hhk}, di-hadron production~\cite{Majumder:2004pt,Zhang:2007ja,Renk:2008xq}, $\gamma$-hadron correlations~\cite{Zhang:2009rn,Qin:2009bk,Wang:2013cia,Chen:2017zte,Luo:2018pto}, reconstructed jets~\cite{Qin:2010mn,Dai:2012am,Wang:2013cia,Chang:2016gjp,He:2018xjv, Qiu:2019sfj, NLLSCET}, and jet substructure~\cite{Chien:2016led,Mehtar-Tani:2016aco,Chang:2017gkt,CasalderreySolana:2012ef,Caucal:2019uvr}. 

These formalisms are based on various approximations that are applicable in limited ranges of shower energy and virtuality scales. Several of these formalisms have recently been implemented in a unified analysis framework, called JETSCAPE~\cite{Cao:2017zih}, providing a multi-stage model of jet evolution in which each  jet quenching formalism is applied only in its appropriate range of validity in shower energy and virtuality. 

Models of in-medium jet-thermal parton interactions have parameters, known as jet transport coefficients, that can be determined by comparison of their calculations to jet quenching measurements~\cite{Gyulassy:1993hr,Baier:1996kr,Baier:1996sk,Baier:1998kq,Zakharov:1996fv,Gyulassy:1999zd,Wiedemann:2000za,Guo:2000nz,Arnold:2002ja,Armesto:2003jh,Djordjevic:2008iz,Majumder:2009ge,CaronHuot:2010bp,Burke:2013yra,CasalderreySolana:2014bpa, Chien:2015vja, Andres:2016iys, NoronhaHostler:2016eow, Bianchi:2017wpt, Zigic:2018ovr}. Phenomenologically, the most significant transport coefficient is \qhat, which denotes the mean square of the momentum transfer between the propagating hard jet and the soft medium per unit length, $\qhat\equiv d\langle k_\perp^2\rangle/dL$, where $\langle\dots\rangle$ indicates the average over all jet propagation paths for the event population.

A quantitative determination of \qhat\ and related quantities, by comparison of theory models with experimental data, has been carried out by several groups~\cite{Burke:2013yra,CasalderreySolana:2014bpa, Chien:2015vja, Andres:2016iys, NoronhaHostler:2016eow, Bianchi:2017wpt, Zigic:2018ovr,Ru:2019qvz,Xie:2019oxg,Xie:2020zdb}. The JET Collaboration extracted \qhat\ from the comparison of multiple jet quenching model calculations to  inclusive hadron  \RAA\ measurements at RHIC and the LHC~\cite{Burke:2013yra}, which are expected to generate a QGP with different initial temperature. Non-perturbative contributions to the value of \qhat\ can also be calculated using first-principles lattice QCD~\cite{Majumder:2012sh} 
though challenges remain to include quark dynamics into a full QGP calculation. A recent 2+1 flavor calculation with physical quark masses on $N_{\tau}$=8 lattices~\cite{Kumar:2019aop,Kumar:2020wvb} 
yields a value for $\qhat/T^3$ in the range of 2.5--3.5 for the highest temperatures generated in heavy-ion collisions at RHIC and the LHC, consistent with the results reported by the JET Collaboration.

Each model in the JET calculation of \qhat\ has a single free parameter, either \qhat\ or the effective strong coupling parameter  \alphas, which is determined separately for RHIC and LHC data. 
In order to go beyond separate extractions of  \qhat\ at RHIC and LHC, and instead obtain a distribution for \qhat\ that is a smooth function of the medium temperature and jet energy, parameter extraction incorporating data from both colliders is required. However, that approach is beyond the scope of the least-squares minimization approach used by the JET analysis, 
and a more comprehensive approach is needed, based on Bayesian inference~\cite{Sacks_1989,Currin_1991,Kennedy_2001}. 

Bayesian inference has been utilized previously to extract parameters of the QGP from heavy-ion collision data, in particular the specific shear viscosity $\eta/s$~\cite{Novak:2013bqa,Bernhard:2016tnd,Everett:2020yty,Everett:2020xug} and the Equation of State (EoS)~\cite{Pratt:2015zsa}, with the latter result agreeing well with Lattice QCD calculations.
See also~\cite{Nijs:2020roc}. This approach has likewise been used to study the heavy quark diffusion coefficient of the QGP~\cite{Xu:2017obm}.

The theoretical description of jet modification includes loss of energy-momentum by hard partons in the shower, generation of softer partons by medium excitation, and excitation of the medium due to energy and momentum exchanges at a scale below that describable by perturbative techniques. A complete description of the entire jet, in terms of all available jet observables, requires the modeling of several transport coefficients. Focus on high-\pT\ hadrons constrains discussion to the hardest partons in the shower, whose distribution is modified by (mainly transverse) momentum exchanges with the medium that depends primarily on \qhat. 

This manuscript presents a quantitative extraction of the temperature and momentum dependence of  \qhat\ in the QGP, using Bayesian inference methods in the JETSCAPE framework. The analysis extends that of the JET collaboration ~\cite{Burke:2013yra} by determining the functional dependence of \qhat\ on the jet energy and virtuality, and the local temperature. Comparison of model calculations with data from both RHIC and the LHC provides a broad scan in jet energy and virtuality.  Two collision centralities are utilized, providing variation in the medium temperature profile.

As the jet shower propagates through the QGP, it exchanges energy and momentum with the dense medium. Momentum exchanges above a certain scale are described using perturbative QCD (pQCD), whereas softer momentum exchanges are modeled by partial thermalization of the exchanged four-momentum with a hydrodynamic background. This calculation incorporates multi-stage jet evolution, using the \Matter\ model~\cite{Majumder:2013re,Cao:2017qpx} at high virtuality scale, and the \Lbt\ model ~\cite{Cao:2017hhk,Cao:2017zih,Chen:2017zte,Luo:2018pto} at low virtuality scale, with the switching between models governed by a free parameter \Qzero. Two different analytic parametrizations of \qhat\ are explored which are functions of the medium temperature and either jet energy or jet virtuality, and which are based on perturbative treatment of jet-medium interactions. 
 
The experimental data used to for Bayesian parameter extraction are measurements of inclusive hadron suppression in central and semi-central \AuAu\ collisions at \sqrtsNN=200 GeV for $\pTh>8$ \gev\ ~\cite{Adare:2012wg}, and \PbPb\ collisions at \sqrtsNN=2.76 and 5.02 TeV for $\pTh>10$ \gev\ ~\cite{Aad:2015wga,Khachatryan:2016odn}. 

The manuscript is organized as follows: Sect.~\ref{sec:jet} presents the jet evolution models; 
Sect.~\ref{sec:qhatParametrization} presents the \qhat\ parametrizations; 
Sect.~\ref{sec:ExpData} presents  the experimental data and treatment of their uncertainties;  
Sect.~\ref{sec:emulator} presents training of the Gaussian process emulator; Sect.~\ref{sec:bayesian} presents implementation of Bayesian inference; Sect.~\ref{sec:Tests} presents closure tests; Sect.~\ref{sec:results} presents the results in terms of posterior distributions for \qhat\ from the Bayesian parameter extraction; and Sect.~\ref{sec:summary} gives a summary and outlook.


\section{Modeling jet-medium interactions} 
\label{sec:jet}

JETSCAPE provides a general numerical framework for simulating jet-medium interactions, with several alternative models to simulate the QGP and jet evolution. QGP evolution is modeled using relativistic hydrodynamics. In this study,  jet evolution is calculated using the \Matter\ model to describe interactions with the QGP at high virtuality and the \Lbt\ model to describe interactions with the QGP at low virtuality. \Matter\ and \Lbt\ are also combined in a multi-stage approach. This section discusses each model in turn.

The primary goal of this calculation is to explore the application of Bayesian inference to the determination of \qhat, with careful assessment of experimental uncertainties. For clarity we therefore limit the complexity of the calculation in its other aspects. The theoretical approach we utilize does not account for all known factors in  the modeling of the  jet-medium interaction and the hydrodynamic medium. The effects we neglect include variation in equilibration time and initial conditions for hydrodynamic evolution~\cite{Andres:2016iys}, the role of event-by-event fluctuations in initial energy density for the hydrodynamic evolution~\cite{NoronhaHostler:2016eow}, and variation in jet quenching model parameters such as start and stop time, and length dependence, of the interaction~\cite{Huss:2020dwe,Huss:2020whe}. We assess the impact of such choices in the following sections. Other calculations addressing the extraction of \qhat\ from inclusive measurements, that each include some of these effects, are found in~\cite{CasalderreySolana:2014bpa,Bianchi:2017wpt,Chien:2015vja}.


\subsection{QGP evolution}
\label{subsec:bulk}

The evolution of the QGP is simulated with second-order relativistic hydrodynamics as implemented in VISH2+1~\cite{Song:2007fn, Song:2007ux, Song:2010mg, Shen:2014vra, Gursoy:2018yai}. An initial entropy density profile from the Monte-Carlo Glauber model~\cite{Miller:2007ri,Shen:link2MCGlauber} is evolved with dissipative fluid dynamics, starting at longitudinal proper time $\tau_0=0.6$\,fm/$c$. For Au-Au collisions at $\sqrt{s_\mathrm{NN}} = 200$~GeV and Pb-Pb collisions at $\sqrt{s_\mathrm{NN}} = 2.76$~TeV, a partial chemical equilibrium equation of state was used with $T_\mathrm{chem}=165$ MeV (s95p-v0-PCE165~\cite{Shen:2010uy,Huovinen:2009yb}), along with a constant  specific shear viscosity $\eta/s=0.08$ and no bulk viscosity; these choices were found to provide a good description of data at RHIC and the LHC in Ref.~\cite{Qiu:2011hf}.  For Pb-Pb collisions at $\sqrt{s_\mathrm{NN}} = 5.02$~TeV, the hydrodynamic profiles were later tuned in Ref.~\cite{Gursoy:2018yai} with a different equation of state and shear viscosity: partial chemical equilibrium equation of state with $T_\mathrm{chem}=150$ MeV (s95p-v1-PCE150~\cite{Shen:2010uy,Huovinen:2009yb}) and a temperature dependent $\eta/s$~\cite{Gursoy:2018yai}. The hydrodynamic model provides the space-time evolution profile of the temperature $T$ and flow velocity $u^\mu$; the viscous part of the energy-momentum tensor is not used. The in-medium jet evolution equations are solved using these profiles, as discussed in the following subsections.

In this work, we apply event-averaged hydrodynamic profiles for jet evolution and only consider jet energy loss inside the QGP phase; interactions in both the brief pre-hydrodynamic stage ($\tau<0.6$\,fm/$c$) and the dilute hadronic stage ($T<T_\mathrm{stop}$) are neglected. The choice of the bulk parameters $\tau_0$, $T_\mathrm{stop}$ and $\eta/s$ affect the calculated value of inclusive hadron \RAA. In the present study their values are fixed by comparing the VISHNU hydrodynamic model with soft hadron data.
Jet-medium interactions are stopped at $T_\mathrm{stop}=165$~MeV, close the the pseudo-critical chiral temperature $T_\mathrm{c}$. If a lower value of $T_\mathrm{stop}$ is used, the jet energy loss will increase and a smaller value of $\hat{q}$ will be extracted from the jet quenching data, though the effect is expected to be small due to the minor enhancement of jet energy loss at such low temperature.


\subsection{Jet production}
\label{subsec:jet production}

Energetic jets are generated in hard scatterings, with rate based on a leading-order perturbative QCD (LO pQCD) calculation in momentum space using the CTEQ5 parametrization of parton distribution functions~\cite{Lai:1999wy}. In nucleus-nucleus collisions, the EPS09 parametrizations of nuclear PDF modification~\cite{Eskola:2009uj} are taken into account for the momentum space distribution of hard partons. Their position space distribution is sampled according to the Monte-Carlo (MC) Glauber model. 


\subsection{Jet-medium interaction at high virtuality: MATTER}
\label{subsec:MATTER}

The {\bf M}odular {\bf A}ll {\bf T}wist {\bf T}ransverse-scattering {\bf E}lastic-drag and {\bf R}adiation (\Matter) model~\cite{Majumder:2013re,Kordell:2017hmi,Cao:2017qpx}  simulates the splitting of highly virtual partons, i.e. jet partons whose virtuality is much larger than the multiple-scattering scale of the medium it probes ($\sim \sqrt{\qhat{E}}$), where $E$ is the parton energy. At high virtuality, the number of splittings dominates over the number of scatterings inside the medium, and the parton splitting process is described by a medium-modified virtuality-ordered shower~\cite{Majumder:2009ge,Majumder:2009zu,Wang:2001ifa,Majumder:2011uk}, where the scattering in the medium provides an additional contribution to the splitting functions. 

A jet shower is initiated by a single hard parton produced at a point $r$ with a forward light-cone 
momentum $p^{+} = (p^0 + \hat{n}\cdot \vec{p} )/\sqrt{2}$, in which $\hat{n} = \vec{p}/| \vec{p} |$ specifies the direction of the jet. The virtuality ($Q$) of a particular parton is sampled based on the Sudakov form factor that determines the virtuality distribution~\cite{Majumder:2013re,Majumder:2014gda}, 

\begin{align}
\label{eq:matter}
\Delta(&Q^2,Q^2_{0})=\prod_a \Delta_a(Q^2,Q^2_{0}) \nonumber\\
&=\prod_a \exp \left[- \int\limits_{Q^2_{0}}^{Q^2}  \frac{dQ^{2}}{Q^{2}} \frac{\alpha_\mathrm{s} (Q^{2})}{2\pi} \int\limits_{z_\mathrm{c}}^{1- z_\mathrm{c}}  dz P_a(z, Q^2) \right].
\end{align}

\noindent
Here, $a$ represents the channels through which the jet parton can split, $Q$ varies from the maximum possible value $Q_\mathrm{max}$ that initiates at the parton energy down to the minimum allowed value of \Qzero\, below which the virtuality-order parton shower breaks.  In the equation above, $z_\mathrm{c}$ is taken as $Q^2_0/Q^2$, and the splitting function contains both vacuum and medium-induced contributions,

\begin{equation}
\label{eq:totP}
P_a(z,Q^2)=P_a^\mathrm{vac}(z)+P_a^\mathrm{med}(z,Q^2).
\end{equation}

\noindent
Here, the medium-induced part is adopted from the higher-twist formalism~\cite{Guo:2000nz,Majumder:2009ge,Aurenche:2008hm,Aurenche:2008mq} and treated as a perturbation to the vacuum part:  

\begin{align}
\label{eq:medP}
P_a^\mathrm{med}&(z,Q^2)=\frac{P_a^\mathrm{vac}(z)}{z(1-z)Q^2}\int\limits_0^\mathrm{\zeta_\mathrm{max}^+}d\zeta^+ \qhat_g(r+\zeta) \nonumber \\
&\times\Biggl[2-2\cos \left(\frac{\zeta^+}{\tau_f^+}\right)-2\frac{\zeta^+}{\tau_f^+}\sin\left(\frac{\zeta^+}{\tau_f^+}\right)\nonumber\\
&+2\left(\frac{\zeta^+}{\tau_f^+}\right)^2\cos \left(\frac{\zeta^+}{\tau_f^+}\right)\Biggl].
\end{align}

\noindent
Here \qhatg\ denotes the gluon transport coefficient; it is evaluated locally at $\vec{r} + \hat{n} \zeta^{+}$ and is related to the quark transport coefficient $\qhat_q$ by color factors. The maximum length sampled $\zeta^+_{\mathrm{MAX}}$ is taken as $1.5 \tau_f^+$, where  $\tau^+_f = 2 p^+/Q^2$ is the mean light-cone formation time. 

After $Q$ of the parent parton is determined, $z$ is chosen by sampling the splitting function $P(z)$. The maximum possible virtualities of the two daughters are thus $zQ$ and $(1-z)Q$, from which the virtualities of the two daughters $Q_1$ and $Q_2$ are similarly assigned by sampling the form factor in Eq.~(\ref{eq:matter}). The transverse (to $\hat{n}$) momentum of the produced pair is then calculated according to the difference in invariant mass between the parent and daughters:

\begin{equation}
\label{eq:kT}
k_\perp^2=z(1-z)Q^2-(1-z)Q^2_1-zQ^2_2.
\end{equation}

The actual time/length for each splitting process is sampled using a Gaussian distribution with a mean value of $\tau^+_f$~\cite{Cao:2017qpx}. This process is iterated until virtualities of all partons within the jet shower reaches the predetermined value of \Qzero. This virtuality-ordered parton shower method is similar to the time-like shower implemented in \textsc{Pythia}, except that here the medium effect is included in a consistent way.

Hard partons evolve through multiple splittings in \Matter\ starting with maximum possible virtualities ($Q=E$) until their virtualities drop below \Qzero. When the \Matter\ model is applied alone, \Qzero\ is fixed at 1~GeV. For proton-proton collisions, only the vacuum contribution to the splitting function Eq.~(\ref{eq:totP}) is taken into account. 
As shown in Ref.~\cite{Cao:2017qpx}, this approach provides a good description of the single inclusive hadron and jet spectra at high \pT\ in proton-proton collisions, serving as a reliable baseline for studying their nuclear modification in heavy-ion collisions. For nucleus-nucleus collisions, both the vacuum and medium-induced parts are implemented. At $Q_0 = 1$~GeV, all partons are converted into hadrons using \textsc{Pythia} fragmentation. 

Partons are fragmented independently using the \textsc{py1ent} function of PYTHIA~\cite{Sjostrand:2006za}. We note that there is sizable uncertainty in parametrized fragmentation functions at LHC energies~\cite{dEnterria:2013sgr}. However, since the combined calculation of initial production and hadronization in JETSCAPE accurately describes jet spectra in proton-proton collisions~\cite{Kumar:2019bvr}, we assume that is can also be used to calculate in-medium modification in heavy-ion collisions.

For the medium-induced part of the splitting function in Eq. (2), the local fluid velocity
of the dynamical medium is taken into account by rescaling \qhat\  in Eq.~(\ref{eq:medP}) via $\qhat=\qhat_\mathrm{local}\cdot p^\mu u_\mu/p^0$~\cite{Baier:2006pt}, where $p^\mu$ is the four-momentum of the jet. The value of \qhat\ is zero before jets enter the thermal medium ($\tau_0<0.6$~fm) and after they exit the QGP; in both regions only the vacuum splitting function contributes to the parton shower. In the remainder of this paper, in the interest of brevity, we will refer to $\qhat_\mathrm{local}$ as $\qhat$. It should be understood that a boost is invoked within the calculation for the case of a moving frame. 

The jet transport coefficient of the QGP medium is the sole parameter of the \Matter\ model. As discussed in~\cite{Cao:2017qpx}, the minimal assumption that \qhat\ is proportional to the entropy density $s$ in the local rest frame, $\hat{q}/s = \hat{q}_0/s_0$, is able to describe single inclusive hadron and jet \RAA, but distinct values of $\qhat_0$ at a given reference point $s_0$ are required at RHIC and the LHC. The present work explores a more general form of \qhat\ as a function of medium temperature, jet energy, and virtuality scale, which can be uniformly applied to data from RHIC and the LHC. 


\subsection{Jet-medium interactions at low virtuality: LBT}
\label{subsec:LBT}

The Linear Boltzmann Transport (\Lbt) model calculates the time evolution of jets in a thermal QGP generated in relativistic heavy-ion collisions, by using a kinetic approach that includes elastic and inelastic collisions~\cite{Wang:2013cia,Cao:2016gvr,Cao:2017hhk,Chen:2017zte,Luo:2018pto,He:2018xjv}. The evolution of the phase space distribution of a jet parton $a$ with $p_a^\mu = (E_a, \vec{p}_a)$ is described using the Boltzmann equation


\begin{equation}
  \label{eq:boltzmann1}
  p_a\cdot\partial f_a(x_a,p_a)=E_a (\mathcal{C}_a^\mathrm{el}+\mathcal{C}_a^\mathrm{inel}),
\end{equation}

\noindent
where $\mathcal{C}_a^\mathrm{el}$ and $\mathcal{C}_a^\mathrm{inel}$ are the collision integrals for elastic and inelastic scatterings.

For elastic scattering of $a$ with a thermal parton $b$ from the medium background, the collision term $\mathcal{C}_a^\mathrm{el}$ is evaluated with the leading-order vacuum matrix elements for all possible $ab\rightarrow cd$ channels. The collinear  ($u,t\rightarrow 0$) divergence of the matrix element is regulated by imposing $S_2(s,t,u)=\theta(s\ge2\mu_\mathrm{D}^2)\theta(-s+\mu_\mathrm{D}^2\le t\le -\mu_\mathrm{D}^2)$, in which $\mu_\mathrm{D}^2=g^2T^2(N_c+N_f/2)/3$ is the Debye screening mass. Therefore, the elastic scattering rate of parton $a$ reads:

\begin{align}
 \label{eq:rate2}
 \Gamma_a^\mathrm{el}&=\sum_{b,c,d}\frac{\gamma_b}{2E_a}\int \prod_{i=b,c,d}d[p_i] f_b(\vec{p}_b) S_2(s,t,u)\nonumber\\
&\times (2\pi)^4\delta^{(4)}(p_a+p_b-p_c-p_d)|\mathcal{M}_{ab\rightarrow cd}|^2,
\end{align}

\noindent
where  $d[p_i]=d^3p_i/[2E_i(2\pi)^3]$, and $\gamma_b$ and $f_b$ represent the spin-color degeneracy and thermal distribution of parton $b$ respectively. The probability of elastic scattering of parton $a$ in each time step $\Delta t$ is thus $P_a^\mathrm{el}=\Gamma_a^\mathrm{el}\Delta t$. 

During the propagation of a jet through the QGP medium, each elastic scattering changes its transverse momentum (perpendicular to its initial direction). This results in an increase of the average transverse momentum square (or transverse momentum broadening) over time. This transverse momentum broadening per unit time (or length) is known as the jet transport coefficient \qhat, characterizing both the local thermal parton density and the strength of jet-medium interaction.  In LBT the value of \qhat\ may be inferred by evaluating Eq.~(\ref{eq:rate2}) weighted by the transverse momentum broadening of the jet parton:

\begin{align}
 \label{eq:qhat0}
 \qhat_a =&\sum_{b,c,d}\frac{\gamma_b}{2E_a}\int \prod_{i=b,c,d}d[p_i] \left[\vec{p}_c-(\vec{p}_c\cdot\hat{p}_a)\hat{p}_a\right]^2 f_b(\vec{p}_b)\nonumber\\
&\times S_2(s,t,u) (2\pi)^4\delta^{(4)}(p_a+p_b-p_c-p_d)\nonumber\\
&\times |\mathcal{M}_{ab\rightarrow cd}|^2
\end{align}

In addition to elastic scattering, inelastic scattering which generates medium-induced gluon radiation is also included in the \Lbt\ model. The inelastic scattering rate at a given time $t$ is defined as the average number of emitted gluons from parton $a$ per unit time, and is evaluated as \cite{Cao:2013ita,Cao:2015hia,Cao:2016gvr}

\begin{equation}
 \label{eq:gluonnumber}
 \Gamma_a^\mathrm{inel} (E_a,T,t) = \frac{1}{1+\delta_g^a}\int dzdk_\perp^2 \frac{dN_g^a}{dz dk_\perp^2 dt},
\end{equation}

\noindent
in which the $\delta_g^a$ term is imposed to avoid double counting for the $g\rightarrow gg$ process. The medium-induced gluon spectrum is taken from the higher-twist energy loss formalism \cite{Guo:2000nz,Majumder:2009ge,Zhang:2003wk},

\begin{eqnarray}
\label{eq:gluondistribution}
\frac{dN_g^a}{dz dk_\perp^2 dt}=\frac{2\alpha_\mathrm{s}(k_\perp^2) P^\mathrm{vac}_a(z) k_\perp^4}{\pi (k_\perp^2+x^2 m_a^2)^4}\,\qhat_g\, {\sin}^2\left(\frac{t-t_i}{2\tau_f}\right).
\end{eqnarray}

\noindent
Here, $z$ and $k_\perp$ are the fractional energy and transverse momentum of the emitted gluon with respect to its parent parton $a$, and $P^\mathrm{vac}_a(z)$ is the vacuum splitting function, and \qhatg\ is the gluon transport coefficient. The initial time $t_i$ denotes the production time of the parent parton $a$ from which the gluon is emitted, and $\tau_f={2E_a z(1-z)}/{(k_\perp^2+z^2m_a^2)}$ is the formation time of the radiated gluon with $m_a$ being the mass of the parton. 

In this work, we assume zero mass for light quarks and gluons. Note that the gluon spectrum is proportional to \qhatg, which is related to the medium parameters in \Lbt\ through Eq.~(\ref{eq:qhat0}).
To avoid possible divergence as $z\rightarrow 0$ as well as violation of detailed balance for low momentum partons, a lower cut-off $z_\mathrm{min}=2\pi T/E$ is implemented for the energy of the emitted gluon~\cite{Cao:2013ita}. Note that Eq.~(\ref{eq:gluondistribution}) is consistent with the medium-induced splitting function Eq.~(\ref{eq:medP}) used in \Matter, except that the $(\zeta/\tau_f)$ and $(\zeta/\tau_f)^2$ terms are ignored here in \Lbt. The contribution from these two terms has been discussed in~\cite{Majumder:2013re} and shown to be small when $\zeta \lesssim \tau_f$. Multiple gluon emissions are allowed in each time step $\Delta t$. Different medium-induced gluon emissions are assumed to be independent of each other; their number $n$ is therefore a Poisson distribution with mean as $\langle N_g^a \rangle = \Gamma_a^\mathrm{inel}\Delta t$,

\begin{eqnarray}
\label{eq:possion}
P(n)=\frac{\langle N_g^a\rangle^n}{n!}e^{-\langle N_g^a\rangle^n}.
\end{eqnarray}

\noindent
Thus, the probability of an inelastic scattering process occurring is $P_a^\mathrm{inel}=1-e^{-\langle N_g^a \rangle}$. Interference effects arising from multiple-gluon emission have not been taken into account. Multiple-gluon emission and resummation of multiple scatterings~\cite{Sievert:2019cwq, Mehtar-Tani:2019ygg, Feal:2019xfl, Andres:2020vxs} will be explored in future work.

To combine the elastic and inelastic processes, the total scattering probability is divided into two regions: pure elastic scattering with probability $P_a^\mathrm{el}(1-P_a^\mathrm{inel})$ and inelastic scattering with probability $P_a^\mathrm{inel}$. The total scattering probability is thus $P_a^\mathrm{tot}=P_a^\mathrm{el}+P_a^\mathrm{inel}-P_a^\mathrm{el}\cdot P_a^\mathrm{inel}$. Based on these probabilities, the Monte Carlo approach is used to determine whether a given jet parton $a$ scatters in the thermal medium, and whether the scattering is purely elastic or inelastic. For a selected scattering channel, the energy and momentum of the outgoing partons are sampled using the corresponding differential spectra given by Eq.~(\ref{eq:rate2}) and (\ref{eq:gluondistribution}). 

For realistic nuclear collisions, we initialize jet partons from hard scatterings in the same way as for the \Matter\ model  (Sect.~\ref{subsec:MATTER}). To account  for the effect of medium flow effect on jet transport during the QGP phase, in each time step we first boost each jet parton into the local rest frame of the fluid cell in which its energy and momentum are updated based on the \Lbt\ model, and then boost it back to the global collision frame where it propagates to the spacetime of the next time step. Note that the previous rescaling of \qhat\ in \Matter\ ($p^0 \hat{q} = (p\cdot u) \hat{q}_\mathrm{local}$) has the same effect as the boost method here when the medium-induced gluon spectrum is written in a boost-invariant form. On the freeze-out hypersurface of the QGP ($T_\mathrm{stop}=165$~MeV), high \pT\ jet partons are passed to \textsc{Pythia 6} for conversion into hadrons. 

In the original work of \Lbt\, the strong coupling constant \alphas\ was the sole parameter determining both the elastic [Eq.~(\ref{eq:rate2})] and inelastic [Eq.~(\ref{eq:gluonnumber})] scattering processes. In this work we take an alternative approach, parametrizing \qhat\ directly (Sect.~\ref{sec:qhatParametrization}).


\subsection{Multi-stage evolution with MATTER+LBT}
\label{subsec:multistage}

Both the medium-modified virtuality shower with \Matter\ alone, and the vacuum virtuality shower with \Lbt\ low-virtuality parton transport, can be used to describe the modification of inclusive hadron and jet distributions in heavy-ion collisions. However, the application of either of these models alone to the entire jet evolution is not theoretically complete; the \Matter\ formalism is not applicable for parton virtuality below the medium scale, and \Lbt\ ignores the in-medium modification of jets in the highly virtual stage. JETSCAPE has therefore developed a multi-stage approach to calculating in-medium jet evolution, in which \Matter\ is applied for partons with high virtuality and \Lbt\ is applied for partons with low virtuality~\cite{Cao:2017zih}. 

For an energetic parton generated by a hard scattering, \Matter\ is used to simulate its virtuality-ordered splitting process (Sect.~\ref{subsec:MATTER}). In each splitting, the virtuality of each of the daughter partons is smaller than that of the parent. When the virtuality of a parton in the shower falls below a specified scale \Qzero, it is passed to \Lbt\ for the subsequent time-ordered in-medium evolution. In this combined approach, \Matter\ largely determines the spectrum of final state partons for high-energy jets or short in-medium path length, while \Lbt\ predominantly governs low energy parton scattering, especially when the in-medium path length is large~\cite{Cao:2017zih}. 

A key parameter of this multi-stage approach is the separation (or switching) scale \Qzero\ between \Matter\ and \Lbt\ evolution, whose value is expected to be similar to that of the medium scale, $Q^2_0 \sim \qhat\tau_f$. A similar separation scale was explored in~\cite{Caucal:2018dla}. Substituting $\tau_f = 2E/Q_0^2$, one obtains $Q_0^2 \sim \sqrt{\hat{q} E} $, as mentioned above. Note that $E$ is the energy of a given parton; in the current implementation, where we transition from one module to another, \Qzero\ has to be replaced by an average scale. While the value of \Qzero\ may be 
considered as a matching scale between the high and low virtuality phases, there is no external physical observable (e.g. jet or hadron \pT) with which it can be linked. We therefore introduce it as a parameter in the Bayesian analysis and 
obtain a mean value and range over which \Qzero\ can be varied. 
This analysis thus provides the first phenomenological determination of \Qzero\ from experimental data, using Bayesian inference.

Note that in the \Lbt\ stage we assume zero virtuality for thermal QGP partons, and recoiling partons are scattered out of the background medium by the jet. In contrast, the virtuality of jet partons is fully tracked since it is fed from \Matter. When a parton splits in \Lbt, we assume that its two daughters share its virtuality in proportion to their $z$ fraction. At temperature below $T_\mathrm{stop} \sim T_\mathrm{c}$, \Lbt\ partons with $Q < 1$~GeV are converted into hadrons using \textsc{Pythia 6}, while those with virtualities still above 1~GeV are passed back to \Matter\ for subsequent vacuum showering until all partons satisfy $Q < 1$~GeV, at which point they are converted to hadrons.

In the pre-equilibrium ($\tau_0<0.6$~fm) and the late ($T<T_\mathrm{stop}$) stages, only \Matter\ vacuum shower is applied with the value of $Q_0 = 1$~GeV. This same vacuum shower is used for $p+p$ collisions. It is only in the hydrodynamic stage ($\tau_0>0.6$~fm and $T>T_\mathrm{stop}$) of \aaa\ collisions that the \Matter ($Q>Q_0$) + \Lbt\ ($Q<Q_0$) model is used with a \Qzero\ value greater than 1~GeV. 

To summarize, in this calculation we 
utilize two different implementations of energy loss, which are applicable in complementary regimes of parton virtuality. This requires the introduction of a separation or matching variable, in this case \Qzero. If a JETSCAPE calculation were to
use a completely different set of energy loss modules which transition in some other variable e.g., parton energy $E$, that would require a different matching variable, say $E_0$, whose value could also be determined using the Bayesian framework that we describe here.



\section{\qhat\ parametrization}
\label{sec:qhatParametrization}

As discussed in Sec.~\ref{sec:jet}, the jet transport coefficient \qhat\ is the sole quantity constrained by fitting \Matter\ and \Lbt\ calculations to experimental data. We employ three different parametrizations of \qhat. The form of the parametrization is derived from Eq.~(\ref{eq:qhat0}), which is based on perturbative scattering of a jet parton inside a medium. Assuming a thermal distribution for $f_b$ and taking the small-angle approximation for elastic scattering gives~\cite{Wang:1996yf,Qin:2009gw,Auvinen:2009qm,He:2015pra}
\begin{align}
 \label{eq:qhat}
 \qhat \approx  C_R\frac{42\zeta(3)}{\pi}\alpha_\mathrm{s}^2 T^3 \ln\left(\frac{2 C E T}{4\mu_\mathrm{D}^2}\right),
\end{align}
where $C_R$ is the color factor of the jet parton (4/3 for quark and 3 for gluon), $T$ is the medium temperature, and $C$ is a constant depending on the kinematic cuts implemented in Eq.~(\ref{eq:qhat}). For the \Lbt\ model using a constant $\alpha_\mathrm{s}=0.3$, $C$ is approximately 5.6 for gluons and 5.8 for quarks~\cite{He:2015pra} . In this work \qhat\ refers to the light-quark jet transport coefficient; the gluon jet transport coefficient is obtained by scaling with the relative color factor. 

The parameter \qhat\ defined above characterizes jet transverse momentum broadening due solely to elastic scattering, which is commonly applied for evaluating medium-induced gluon emission. An additional double-logarithmic dependence of jet transverse momentum broadening would arise if radiative processes are taken into account~\cite{Liou:2013qya,Arnold:2020uzm}.

The definition of \qhat\ need not be limited to perturbative scattering of a jet parton with a thermal medium at the scale  $T$. Thus, we extend Eq.~(\ref{eq:qhat}) to a more general form as follows:
\begin{widetext}
\begin{equation}
\label{eq:parametrization1}
\frac{\qhat\left(E,T\right) |_{A,B,C,D}}{T^3}=42C_R\frac{\zeta(3)}{\pi}\left(\frac{4\pi}{9}\right)^2\left\{\frac{A\left[\ln\left(\frac{E}{\Lambda}\right)-\ln(B)\right]}{\left[\ln\left(\frac{E}{\Lambda}\right)\right]^2}+\frac{C\left[\ln\left(\frac{E}{T}\right)-\ln(D)\right]}{\left[\ln\left(\frac{ET}{\Lambda^2}\right)\right]^2}\right\},     
\end{equation}
\end{widetext}
\noindent
where $(A,B,C,D)$ are parameters that will be determined from the experimental data using Bayesian parameter extraction. If the first part in the braces $\{...\}$ (or parameter $A$) is set to zero, the second part reduces to Eq.~(\ref{eq:qhat}) if the coupling constant $\alpha_\mathrm{s}=4 \pi/9/\ln(ET/\Lambda^2)$ is assumed to run with both jet energy and medium temperature scales at leading order. For the parameter $\Lambda$ we use $\Lambda= 0.2$\,GeV.

The first part of the expression in the braces is an ansatz applicable to a highly energetic parton whose virtuality is much higher than the thermal scale of the medium, and which is therefore blind to the thermal scale. In this case, after being scaled by the density of the scattering centers ($\sim{T^3}$), the value of \qhat\ is controlled solely by the scale of the jet parton itself, and not by the medium temperature. This first part in $\{...\}$, with parameters $A$ and $B$, represents the physics assumed by the \Matter\ model. 

The second part in $\{...\}$ represents an on-shell jet parton scattering with quasi-particles inside a thermal medium, as assumed by the \Lbt\ model. The arguments of the logarithms in Eq.~(\ref{eq:parametrization1}) involve additional constant factors that depend on the particular cut-off value implemented in the $t$-channel scattering. We treat these as parameters, called $B$ and $D$, even though the jet observables considered here are not expected to be very sensitive to them. This expectation is validated by their broad posterior distribution function obtained from the model-to-data comparison, as shown below.

We consider Eq.~(\ref{eq:parametrization1}) to be a a sufficiently general ansatz of the energy and momentum dependence of \qhat\ within the perturbative picture of jet-medium interaction. We use this parametrization consistently in both \Matter\ and \Lbt\ when they are applied separately to describe experimental data. We expect that the physics of the high virtuality stage in \Matter\ is described predominantly by the first term (with $A$ and $B$), while the physics of the thermal stage in \Lbt\ is described by the second term (with $C$ and $D$). 

For the multi-stage calculation combining \Matter+\Lbt\ we utilize two different parametrizations of \qhat. The first parametrization uses Eq.~({\ref{eq:parametrization1}) to calculate \qhat\ in both \Matter\ and \Lbt\ stages, while introducing an additional parameter \Qzero\ that represents the virtuality boundary between the two stages. This five-parameter formulation is denoted ``\Matter+\Lbt~1}". Since it is based on the same physical assumptions as \qhat, it can be compared directly to the parametrization in which \Matter\ and \Lbt\ are applied separately.

To reduce the number of parameters and capture the jet physics of virtuality evolution in \Matter\ more precisely, we introduce a second \qhat\ parametrization for the multi-stage \Matter+\Lbt\ model, as follows:

\begin{widetext}
\begin{equation}
\label{eq:parametrization2}
\frac{\qhat\left(Q,E,T\right) |_{\Qzero,A,C,D}}{T^3}=42C_R\frac{\zeta(3)}{\pi}\left(\frac{4\pi}{9}\right)^2\left\{\frac{A\left[\ln\left(\frac{Q}{\Lambda}\right)-\ln\left(\frac{Q_0}{\Lambda}\right)\right]}{\left[\ln\left(\frac{Q}{\Lambda}\right)\right]^2}\theta (Q-Q_0)+\frac{C\left[\ln\left(\frac{E}{T}\right)-\ln(D)\right]}{\left[\ln\left(\frac{ET}{\Lambda^2}\right)\right]^2}\right\}.
\end{equation}
\end{widetext}

\noindent
This parametrization is denoted ``\Matter+\Lbt~2". Compared to Eq.~({\ref{eq:parametrization1}), we use the jet virtuality $Q$ as the scale in the first term instead of the jet energy $E$. 
The motivation behind this parametrization is that the \Matter\ model better characterizes the parton shower as a function of virtuality. However, the value of \qhat\ determined using this parametrization cannot be directly compared to that from \Lbt.

The parameter $B$ is replaced by the switching virtuality \Qzero, so that this formulation likewise has four parameters. The $\theta$ function   ensures that, during the \Matter\ stage ($Q>\Qzero$), \qhat\ receives contributions from both terms, while during the \Lbt\ stage ($Q<\Qzero$) only the second term contributes. In this parametrization, the distribution of \qhat\ is continuous at $Q=\Qzero$. 

\section{Experimental data}
\label{sec:ExpData}

This analysis carries out Bayesian parameter extraction using experimental measurements of inclusive hadron production in \aaa\ collisions at RHIC and LHC (\RAA). Selection of experimental data for this process requires consideration of the \pT\  range suitable for comparison to theoretical calculations of jet quenching, in particular the possible role of medium-modified hadronization at low \pT.

The energy loss formalism in this manuscript involves the convolution of initial state and hard scattering distributions with   energy loss calculations applied to hard partons as they propagate through the medium. The final parton distributions are then convoluted with vacuum fragmentation functions to calculate the hadron distributions to be compared to data. The calculations are therefore based on the assumption that the hadronization of leading hadrons takes place outside the dense medium.

The space-time distribution of jet hadronization in the presence of the QGP is currently an open issue, to be resolved using both experimental data and theoretical modeling. Relevant experimental data to address this question include high-\pT\ di-hadron correlations in central \AuAu\ collisions at RHIC~\cite{Adams:2006yt}, whose jet-like angular distributions indicate that charged hadrons with $\pT\gtrsim{4}$ \gev\ arise predominantly from vacuum fragmentation; and particle-identified relative yields in reconstructed jets in central \PbPb\ collisions at the LHC, which are similar to those for jets in vacuum for $\pT>4$ \gev, in contrast to a striking enhancement in the baryon/meson ratio for bulk (non-jet) production~\cite{Kucera:2015fni}. These experimental observations suggest that hadrons with $\pT\gtrsim{4}$ \gev\ are generated in central \aaa\ collisions predominantly by jet fragmentation in vacuum; in other words, that the processes of jet-medium interactions and hadron formation largely factorize for hadrons with $\pT>4$ \gev. On the other hand, parametric theoretical arguments suggest that vacuum hadronization occurs only for hadrons with $\pT>10$ \gev, at both RHIC and the LHC. 

In order to simplify the analysis presented in this manuscript, we therefore restrict the \pT\ range of the inclusive hadron \RAA\ measurements considered for comparison to the theory calculations to $\pT>8$ \gev\ at RHIC and $\pT>10$ \gev\ at the LHC. We note that this cut limits significantly the statistical weight of the RHIC data relative to that at the LHC, due to the much narrower kinematic range accessible at the lower \sqrtsNN\ of RHIC (see Sect.~\ref{sec:results}). Lowering of this limit, to enable greater statistical weight of RHIC data, will be explored in future work.

The experimental datasets used in this analysis, which cover a wide range in hadron \pT\ and medium temperature, are as follows:

\begin{itemize}
\item Au-Au collisions at $\sqrtsNN=200$~GeV, 0-10\% and 40-50\% centrality~\cite{Adare:2012wg};
\item  Pb-Pb collisions at $\sqrtsNN=2.76$~TeV,  0-5\% and 30-40\% centrality~\cite{Aad:2015wga};
\item  Pb-Pb collisions at $\sqrtsNN=5.02$~TeV,  0-10\% and 30-50\% centrality~\cite{Khachatryan:2016odn}.
\end{itemize}


\subsection{Experimental uncertainties}
\label{subsec:ExpUncert}

Bayesian parameter extraction requires specification of experimental uncertainties, optimally the full covariance matrix $\Sigma_E$. However, the full covariance matrix of measurement uncertainty is difficult to determine, and it is usually not reported in experimental publications. We focus here on measurements of inclusive hadron \RAA, and discuss how the covariance matrix  is determined for the reported measurements used in this analysis.

For these measurements, the experimental uncertainties are specified as a function of hadron \pT. The CMS~\cite{Khachatryan:2016odn} and ATLAS~\cite{Aad:2015wga} publications report the following four types of uncertainty:

\begin{enumerate}[itemsep=0mm]
\item uncorrelated statistical error and systematic uncertainty on each data point; 
\item luminosity uncertainty, fully correlated in all centrality bins for a given collision system;
\item Glauber scaling (\TAAavg)  uncertainty, fully correlated in \pT\ for a given collision system and centrality bin; and
\item other correlated errors of unspecified origin, with only qualitative dependence on hadron \pT\ specified. 
\end{enumerate}

Because the luminosity and \TAAavg\ uncertainties are independent of \pT, it is straightforward to calculate their contribution to the off-diagonal terms of the covariance matrix.  However, the other correlated uncertainties arise from sources such as track selection, momentum resolution, and efficiency correlations, which vary in different ways with \pT.  To account for this complexity we introduce a correlation-length parameter $\ell$ (defined below) to represent the range in \pT\ over which these uncertainties contribute.

For the \RAA\ distribution of a specific collision system and centrality from a specific experimental publication (indexed by $k$), let $\Sigma^E_k$ be the corresponding covariance matrix block constructed from ``uncorrelated'', ``fully-correlated'', and ``length-correlated" uncertainty vectors respectively, $\{\sigma_{k}^\text{uncorr}\}$, $\{\sigma_{k}^\text{fcorr}\}$, $\{\sigma_{k}^\text{lcorr}\}$, as reported by the experiments. Then the uncertainty covariance block $\Sigma^E_k$ is given by

\begin{eqnarray}
&&\Sigma_k^E = \Sigma_k^\text{uncorr} + \Sigma_k^\text{fcorr} + \Sigma_k^\text{lcorr} \nonumber \\
&&\Sigma_{k,ij}^\text{uncorr} =  \sigma_{k,i}^\text{uncorr}\sigma_{k,j}^\text{uncorr}\delta_{ij} \nonumber \\
&&\Sigma_{k,ij}^\text{fcorr} =  \sigma_{k,i}^\text{fcorr}\sigma_{k,j}^\text{fcorr}\nonumber \\
&&\Sigma_{k,ij}^\text{lcorr} = \sigma_{k,i}^\text{lcorr}\sigma_{k,j}^\text{lcorr}\text{exp}\left[-\left|\frac{p_{k,i} - p_{k,j}}{\ell_k}\right|^\alpha\right].
\label{eq:errors}
\end{eqnarray}

\noindent
Here $p_{k,i}$ is the $i^\mathrm{th}$ \pT\ value in block $k$, and $\delta_{ij} = 1$ if $i=j$ and 0 otherwise. Thus, $\Sigma^\text{uncorr}_k$ is a diagonal matrix, representing the combined, uncorrelated statistical and systematic experimental uncertainties.  $\Sigma^\text{fcorr}_k$ corresponds to the fully correlated,  \pT-independent luminosity and \TAAavg\ uncertainties, and $\Sigma^\text{lcorr}_k$ is constructed from the correlated experimental uncertainties using a power exponential covariance function. We set the exponent $\alpha = 1.9$, similar to the common choice $\alpha=2$ but computationally more stable~\cite{Gu_2016}.  The $p_{k,i}$ transverse momentum values and correlation length $\ell_k$ in Eq.~(\ref{eq:errors}) are linearly rescaled so that all values lie within [0,1].  The rescaled correlation length $\ell_k$ is nominally set to a value of $\ell_k=0.2$.  Other values were used to study the sensitivity of our results to this parameter choice.

The PHENIX publication~\cite{Adare:2012wg} reports uncertainties in a similar fashion but with different labels. Uncorrelated errors are denoted as Type A, fully correlated and \pT-independent are reported as Type C, and Type B refers to correlated systematic errors with an unspecified \pT-dependence.  Therefore Type A, B, and C errors are treated as $\sigma^\text{uncorr}$, $\sigma^\text{lcorr}$, $\sigma^\text{fcorr}$, respectively, according to Eq.~(\ref{eq:errors}).


\section{Gaussian process emulators}
\label{sec:emulator}

Because JETSCAPE model calculations are computationally expensive, we use Gaussian Process Emulators (GPEs) to interpolate the model-parameter space~\cite{Sacks_1989,Rasmussen:2006}.  GPEs offer a non-parametric method of regression, providing a statistical surrogate for the computationally expensive model by using a limited set of training points to predict with a defined uncertainty any untried value of input parameters.  This allows us to make rigorous statistical comparisons to the experimental data efficiently, and perform inference on the input parameters.

Our implementation of the GPE is identical to that used in~\cite{Bernhard_2015}, except for the choice of the covariance function which controls the correlation between pairs of points.  To improve emulator stability, we replace the squared-exponential function used in~\cite{Bernhard_2015} with a Mat\'{e}rn 5/2 covariance function,

\begin{equation}
	c(\mathbf{x}_i,\mathbf{x}_j)  = \left(1 + \frac{\sqrt{5}d_{i,j}}{\ell} + \frac{5d_{i,j}^2}{3\ell^2} \right) \exp \left( - \frac{\sqrt{5}d_{i,j}}{\ell} \right)
\label{eq:Matern}
\end{equation} 
	

\noindent
where $d_{i,j} = |x_{i} - x_{j}|$ denotes the difference between pairs of points.
The correlation length parameters \{$\ell$\} are found through hyperparameter optimization in the scikit-learn package \cite{scikit-learn}.

By far the most CPU-intensive part of the present study is calculating the result of the physics model for each design point sample in the parameter space. For each choice of centrality bin and colliding system, we simulate over 10 million jet events  for a given model setup and set of parameters (or training point). A single such simulation requires over 1000 CPU hours. Since multiple colliding systems, model setups and training points are utilized for this study, over 10 million CPU hours have been utilized in total on the Open Science Grid.

%

\subsection{Design points}

\begin{figure}[tbp]
  \begin{center}
   \includegraphics[width=0.5\textwidth]{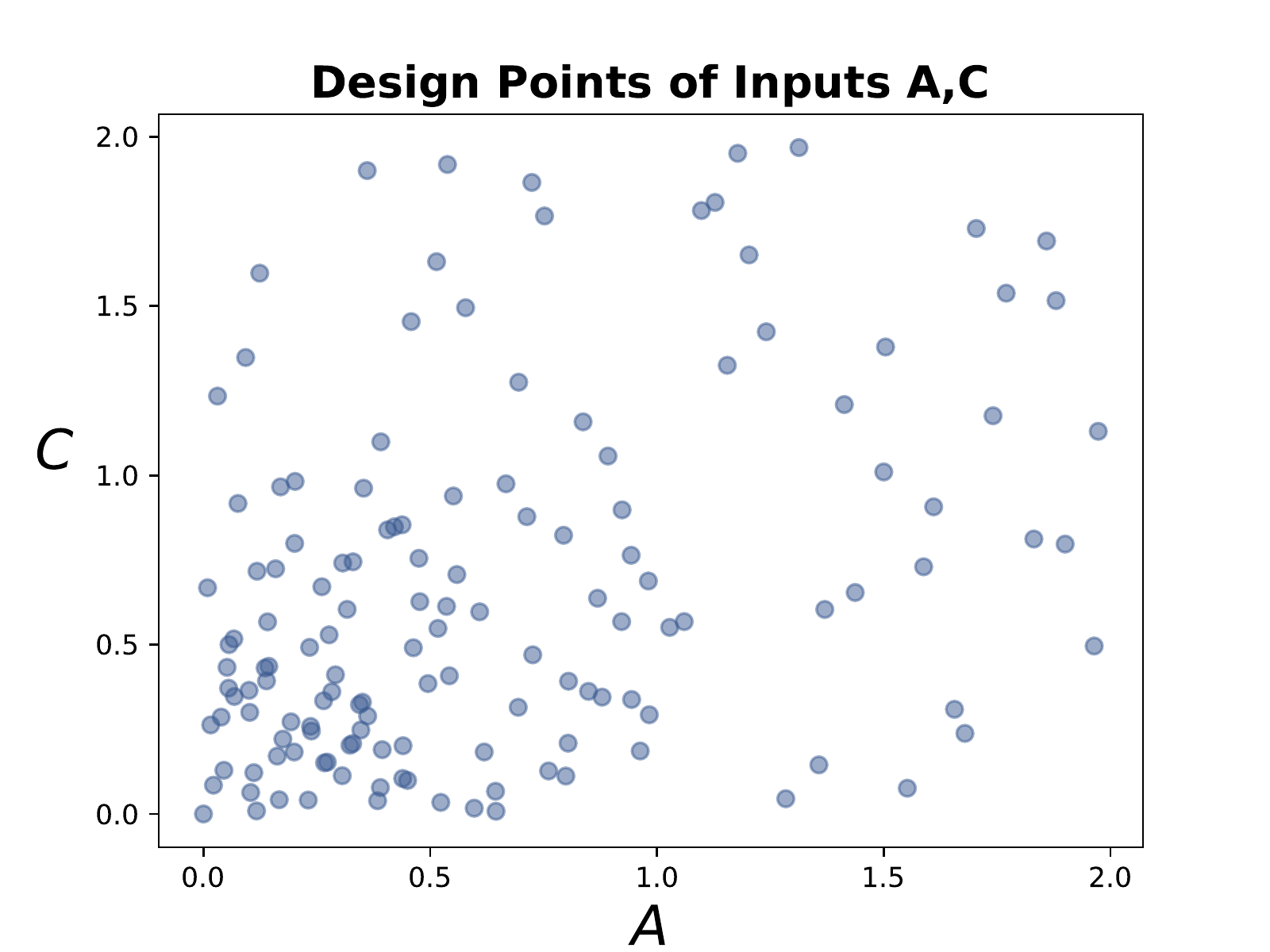}  
  \caption{Latin Hypercube Design for input parameters $A$ and $C$ for the \Lbt\ model.}
  \label{fig:design}
  \end{center}
\end{figure}

Performance of the GPE depends critically on the choice of design points. We base our initial choice of the design points on the method of a space-filling Latin Hypercube Design (LHD) \cite{McKay_1979,Stocki_2005}, which ensures marginal uniformity and optimizes the distance between points.
This was implemented with the function \texttt{optimumLHS} in the \texttt{R} package \texttt{lhs}.
In order to reduce the emulator interpolation uncertainty in the most relevant regions of phase space, the choice is then revisited and improved by adding more design points. For instance, for the \Lbt\ model calibration described later, we start with 60 training points uniformly sampled within $A \times C \in [0, 2]\times [0, 2]$. Based on their preliminary posterior distribution after the entire analysis procedure (as will be discussed in the following sections), we sequentially add more training points -- 20 points in $[0, 0.4] \times [0, 0.4]$, 40 points in $[0, 1] \times [0, 1]$ and then 20 points in $[0, 0.6] \times [0, 0.75]$ -- and repeat the calibration procedure several times to ensure that sufficient training points have been sampled within the region where the peaks of the posterior distributions of our model parameters locate. 
Note that the design point parameter space has four or five dimensions, which cannot be directly visualized. To illustrate the final set of design points, the distribution of inputs $A$ and $C$ for the \Lbt\ model is shown in Fig.~\ref{fig:design}.

\subsection{Multivariate output}

For each design point, we run the computer model for three collision systems and two centralities (Sect.~\ref{sec:ExpData}) to determine the inclusive hadron  \RAA\ at various \pT\ values. The set of \RAA\ values at each \pT\ provides a 66-dimensional output for each design point. Instead of passing the 66-dimensional output directly to a high-dimensional GPE, we first employ a Principal Component Analysis (PCA). The PCA both reduces the output dimensionality and provides a linearly independent description, making the output data more tractable and allowing the application of independent GPEs. 

PCA rotates the data onto an orthogonal space, utilizing the Singular Value Decomposition (SVD) of the data, as follows. Let $\mathbf{Y}$ be the centered and scaled output with $n$ rows and $p$ columns; then for a diagonal matrix $\mathbf{S}$ and orthogonal matrices $\mathbf{U}$ and $\mathbf{V}$, the SVD of $\mathbf{Y}$ is 
\begin{equation}
	\mathbf{Y} = \mathbf{USV}',
\end{equation}

\noindent
where $\mathbf{V}'$ denotes the transpose of $\mathbf{V}$. The rotation of $\mathbf{Y}$ by $\mathbf{V}$ gives the matrix $\mathbf{US}$, which has uncorrelated columns. If we assume normality of the data, then the columns are independent as well. Thus, we apply the transformation

\begin{equation}
	\mathbf{Z} = \mathbf{YV}
\end{equation}

\noindent
and train independent GPs on the columns of $\mathbf{Z}$. To predict a new point, we take the GP predictions and rotate them by $\mathbf{V}'.$

PCA can also be used for dimension reduction. The values $\{s_r\}$ of $\mathbf{S}$ are the square roots of the eigenvalues $\{\lambda_r\}$ of the scaled sample covariance  matrix of $\mathbf{Y}$, in non-increasing order, with associated eigenvectors in $\mathbf{V}$. The fraction of variance corresponding to the first $R$ eigenvectors, for $R \leq p$, is

\begin{equation}
	F_R = \frac{\sum_{r=1}^R\lambda_r}{\sum_r \lambda_r}.
\end{equation}

\noindent
If we use only the first $R$ columns of $\mathbf{V}$ in our transformation (call this matrix $\mathbf{V}_R$), we capture $F_R$ of the variance explained by $\mathbf{Y}$. Thus the goal is to choose a value of $R$ that is large enough to explain a suitable amount of variance, and small enough for a tractable number of GPEs.  Figure~\ref{fig:F_R} depicts the variance corresponding to $R$ for the \Lbt\ data. For these data we choose $R=3$, which captures 99.9\% of the variation. 

\begin{figure}[tbp]
  \begin{center}
  \includegraphics[width=0.5\textwidth]{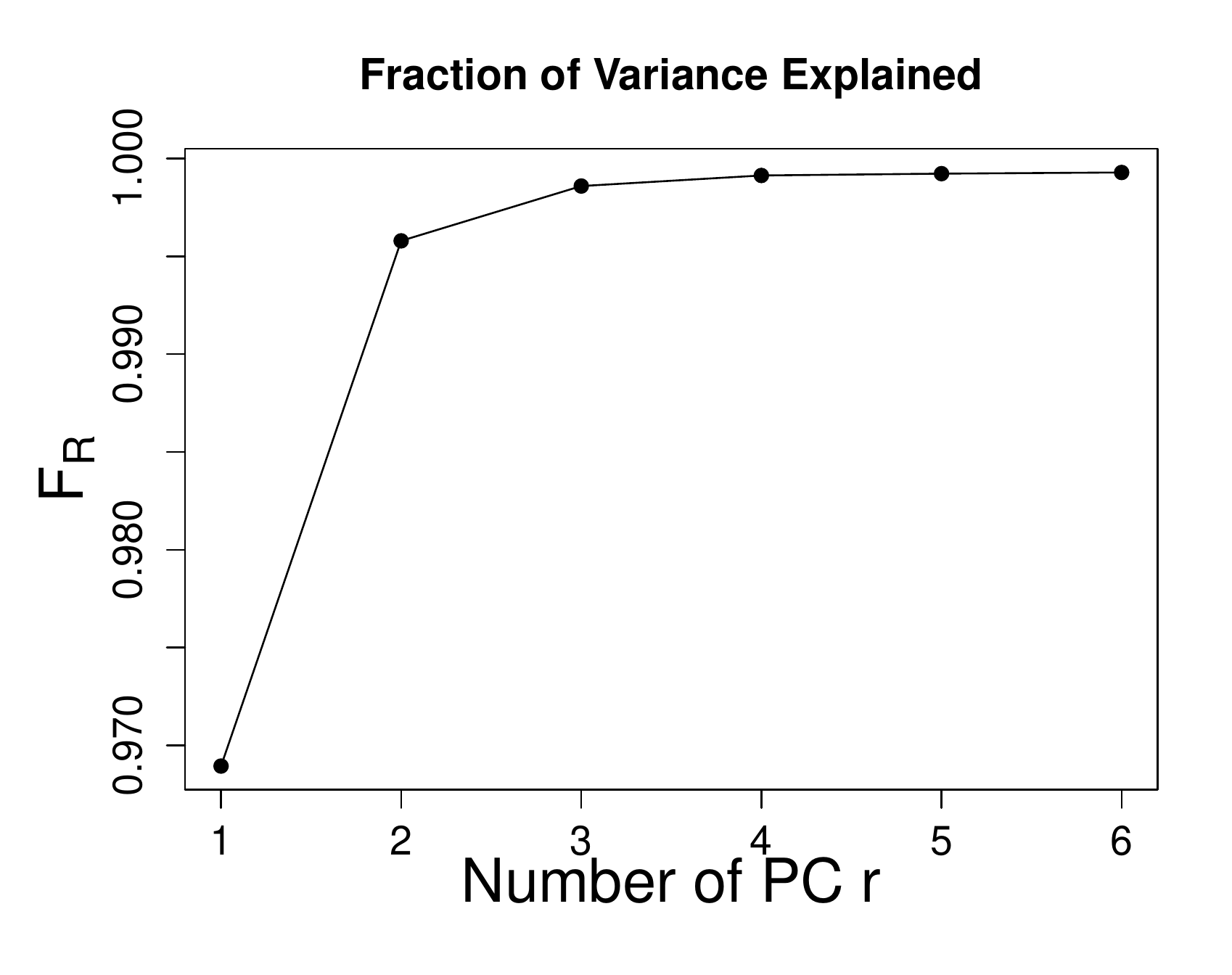}  
  \caption{Percent of the output data variance as a function of the number of components determined in the PCA.  99\% of the variance is described with two components, while 99.9\% is described with three components.}
  \label{fig:F_R}
  \end{center}
\end{figure}

\subsection{Emulator validation}
\label{subsec:emulatorValidation}

To validate our GPEs, we perform ``hold out" tests in which we remove a design point from the emulator training procedure, 
and compare the emulator predictions at that design point to the corresponding model calculation. 
We repeat this procedure for all design points, in order to validate the emulator performance broadly across the design space.
Figure~\ref{fig:emulator} depicts examples of these comparisons for the LBT model, along with a comparison to the emulator uncertainties in the right panel. 
We find that the GPEs generally predict the model well, and that the emulator uncertainties capture the deviations reasonably well;
the emulator uncertainties in fact slightly overestimate the observed deviations.
There exist a small number of design points which are poorly predicted, as shown by the off-diagonal scatter points. 
These points originate at the boundaries of the parameter space, where interpolation is not possible; we verified that they do not impact our results. 
We validated the GPE performance for all models, with the average emulator uncertainties in the range $\left< \sigma_{\rm{emulator}} \right> \approx 8-22\%$ (\Lbt),
$\left< \sigma_{\rm{emulator}} \right> \approx 5-14\%$ (\Matter), $\left< \sigma_{\rm{emulator}} \right> \approx 4-14\%$ (\Matter+\Lbt\ 1), and $\left< \sigma_{\rm{emulator}} \right> \approx 9-24\%$ (\Matter+\Lbt\ 2), depending upon collision system. 

\begin{figure}[tbp]
  \begin{center}
  \includegraphics[width=0.5\textwidth]{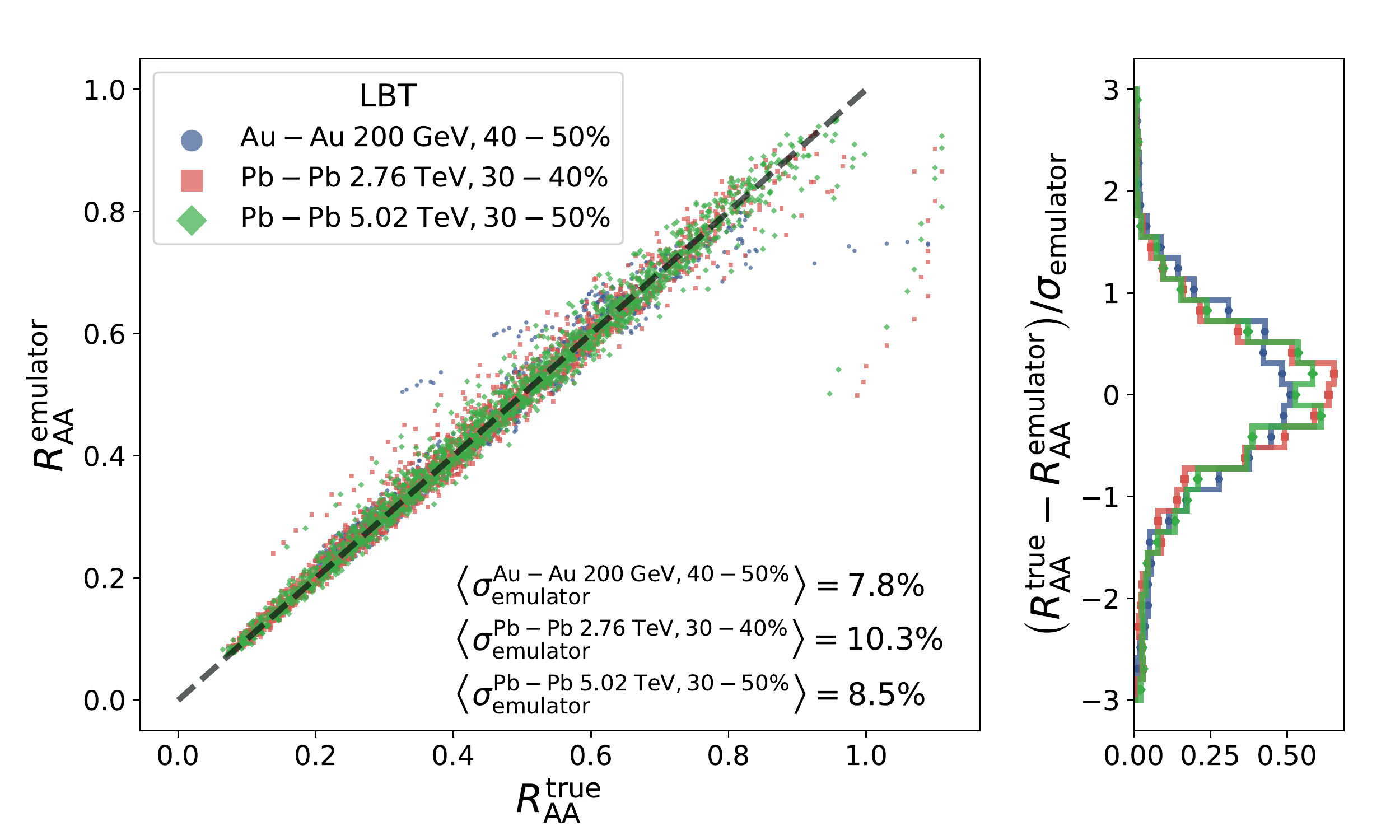}  
  \caption{Validation of the Gaussian process emulator predictions. For each design point, the emulators are re-trained without constraint from that
  holdout point, and the emulator predictions are compared with the model calculations at the design point.}
  \label{fig:emulator}
  \end{center}
\end{figure}


\section{Bayesian calibration}
\label{sec:bayesian}

With a validated emulator, we can proceed to \emph{calibration} - using the experimental data to perform inference on the input parameters. We use Bayesian inference, treating
model parameters as random variables characterized by probability distributions, and use Bayes' Rule to update the prior distribution of input parameters $\thetaE$ (e.g. for LBT, these are $\thetaE=\{A,B,C,D\}$)
to the posterior distribution $\thetaE$ conditional on the experimental values $\mathbf{Y}_E$ \cite{Kennedy2001, BDA3}.
Let $m(\theta)$ denote the computationally expensive computer model; then, the posterior distribution $f(\thetaE\mid \mathbf{Y}_E)$ is

\begin{equation}
	f(\thetaE\mid \mathbf{Y}_E) \propto f(\mathbf{Y}_E\mid m(\thetaE))f(\thetaE).
\end{equation}

\noindent
Because the posterior $f(\thetaE\mid \mathbf{Y}_E)$ is not analytically tractable, we employ an affine invariant Markov Chain Monte Carlo algorithm \cite{emcee} to draw samples from $f(\thetaE\mid \mathbf{Y}_E)$. 

Recall that we train $R$ independent GPEs on the first $R$ columns of $\mathbf{Z = YV}$. Let $m_r^*(\thetaE)$ be the GPE interpolation for given inputs $\thetaE$.  Then $m_r^*(\thetaE)$ has Normal distribution with mean $\mu_r^*(\thetaE)$ and variance ${\sigma_r^*}^2(\thetaE)$.

Note that each predictive mean  $\mu_r^*(\thetaE)$ and variance ${\sigma_k^*}^2(\thetaE)$ are implicitly conditioned on column $r$ of $\mathbf{Z}$ (i.e. transformed design output) and design input. Because the GPEs are independent, we can easily write down the joint distribution of $m^*(\thetaE) = [m_1^*(\thetaE),\ldots,m_R^*(\thetaE)]'$:

\begin{align}
m^*(\thetaE) &\sim \N(\mu^*(\thetaE),\Sigma^*(\thetaE))  \nonumber \\
\mu^*(\thetaE) &= \left[\mu_1^*(\thetaE),\ldots,\mu_R^*(\thetaE)\right]' \nonumber  \\
\Sigma^*(\thetaE) &= \texttt{diag}\left(\left[{\sigma_1^*}^2(\thetaE),\ldots, {\sigma_R^*}^2(\thetaE)\right]'\right)
\end{align}

Since we emulate in PCA space, we must rotate our predictive interpolations back into the observable space, i.e. multiply $m(\thetaE)$ by $V_R'$. However, even though we capture over 99\% of the variance with our choice of $R$, we have found calibration to be more stable if we add back the extra variation lost when transforming back to the physical space.  From the SVD decomposition $\Ybf = \mathbf{USV}'$, we see that $\mathbf{Y'Y} = \Vbf\Sbf^2\Vbf'.$ Additionally, if we let $V_b$ denote the matrix comprised of the columns of $\Vbf$ from $R+1$ onward (and similarly to $\Sbf$) then we can decompose the $\Vbf\Sbf^2\Vbf'$ into the sum

\begin{equation}
\Vbf\Sbf^2\Vbf' = \Vbf_R\Sbf_R^2{\Vbf_R}' +  \Vbf_b\Sbf_b^2{\Vbf_b}'.
\end{equation}

\noindent
Noting that the sample covariance matrix is $\frac{1}{n}\mathbf{Y'Y}$, we denote $\Sigma_\text{extra} = \frac{1}{n}\Vbf_b\Sbf_b^2{\Vbf_b}'$ the covariance matrix of extra variation lost when transforming back and forth from the PCA space.

Initially, we model $\mathbf{Y}_E$ (centered and scaled to match $\mathbf{Y}$) as multivariate Normal, centered at $m^*(\thetaE)V'$ with covariance matrix $\Sigma_E+ \Sigma_\text{extra}$. However, because $m^*(\thetaE)V'$ is also multivariate Normal, we can analytically integrate over $m^*(\thetaE)$. Our final calibration model is thus

\begin{align}
f(\mathbf{Y}_E\mid \thetaE) &\sim \N(\mu^*(\thetaE)\Vbf_R',\Vbf_R\Sigma^*(\thetaE)\Vbf_R' + \Sigma_E + \Sigma_\text{extra})\nonumber \\
   f(\thetaE) &\sim \texttt{unif}(\thetaE),
\end{align}

\noindent
where we assign a uniform prior on the design space for $\thetaE$. 

To sample the posterior distribution, we discard the first 30,000 samples (``burn-in'') of the Markov Chain Monte Carlo algorithm, for which the sampler has not yet reached equilibration, and then save the next 100,000 as draws from the posterior distribution $f(\thetaE\mid \Ybf_E)$. 

\section{Closure tests}
\label{sec:Tests}

\begin{figure}[tbp]
   \centering
    \includegraphics[width=0.9\linewidth]{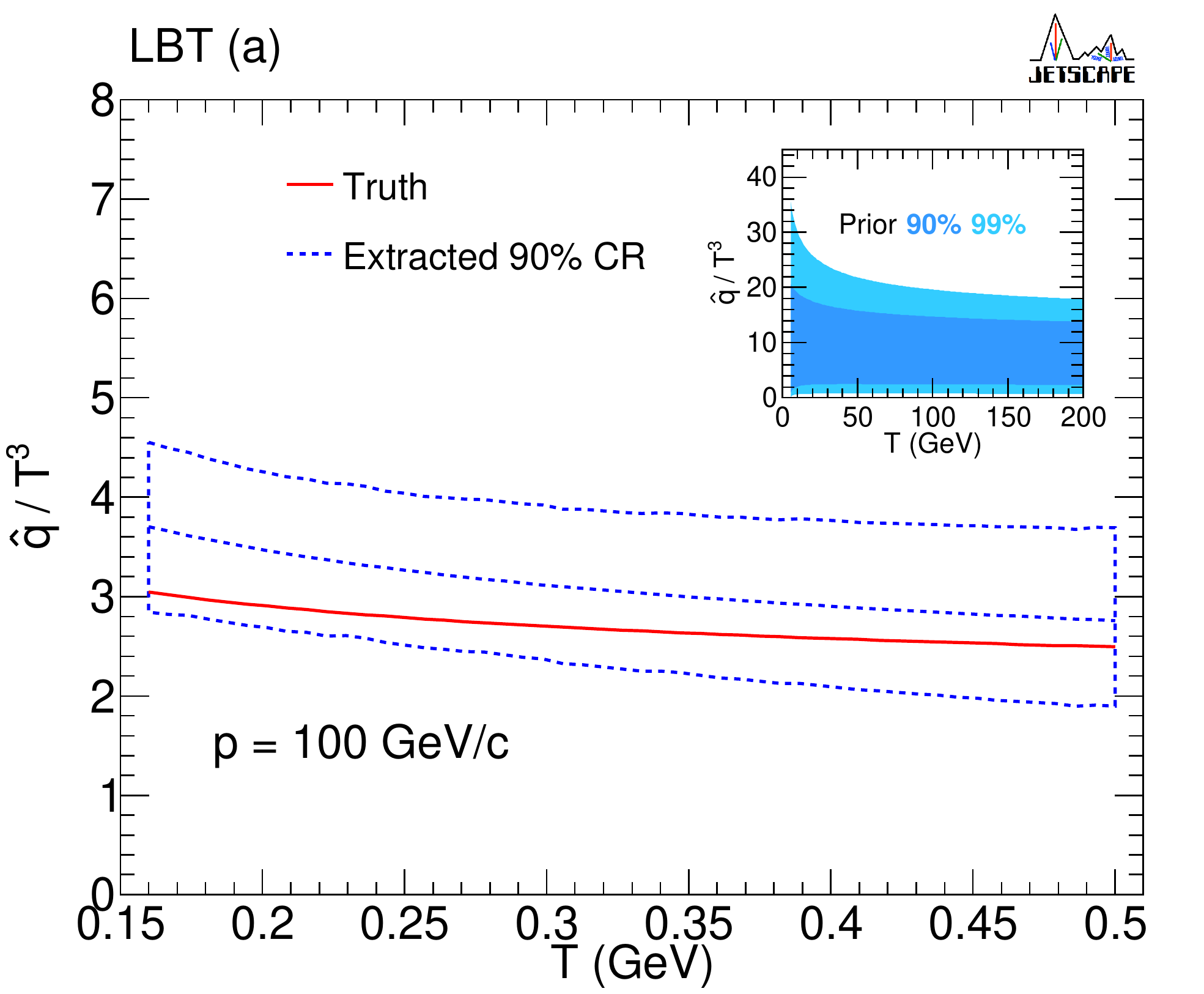}
    \includegraphics[width=0.9\linewidth]{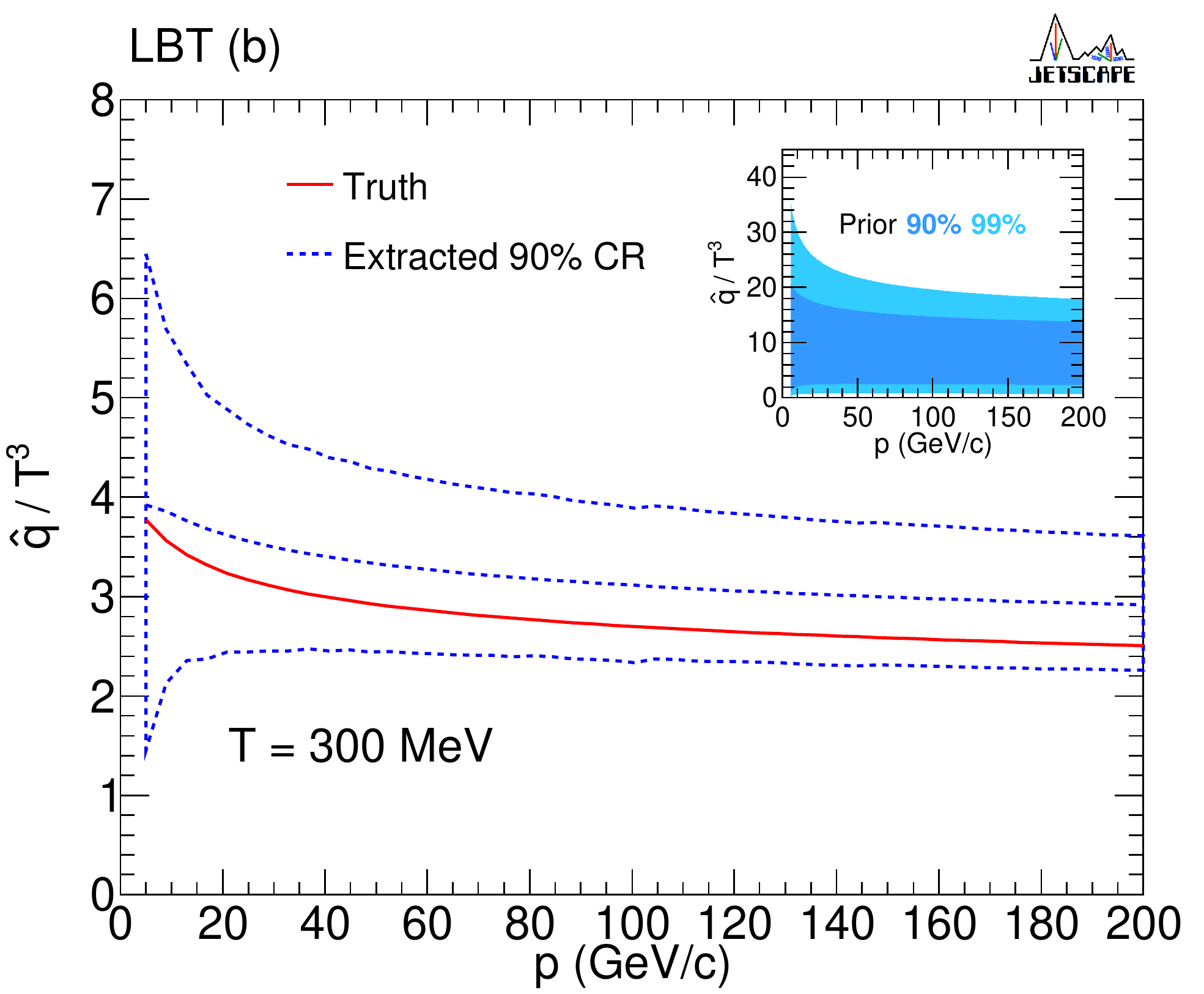}
   \caption{Example closure test for a single design point.  The truth (solid line) is compared to the inferred 90\% range of \qhat\ (band delineated by dashed lines), as a function of (a) temperature and (b) jet momentum.}
   \label{fig:closure-example}
\end{figure}

In order to validate the end-to-end analysis procedure, we perform a set of closure tests. We ``hold out" a design point from the emulator training,
as described in Section \ref{subsec:emulatorValidation}, and instead use the model predictions at that design point to generate ``pseudo-data" equivalent
to the experimentally measured datasets.
We then perform the Bayesian calibration procedure using this pseudo-data in place of the experimental measurements, and compare the inferred parameters to the original parameters of the design point.

\begin{figure}[tbp]
   \centering
   \includegraphics[width=0.9\linewidth]{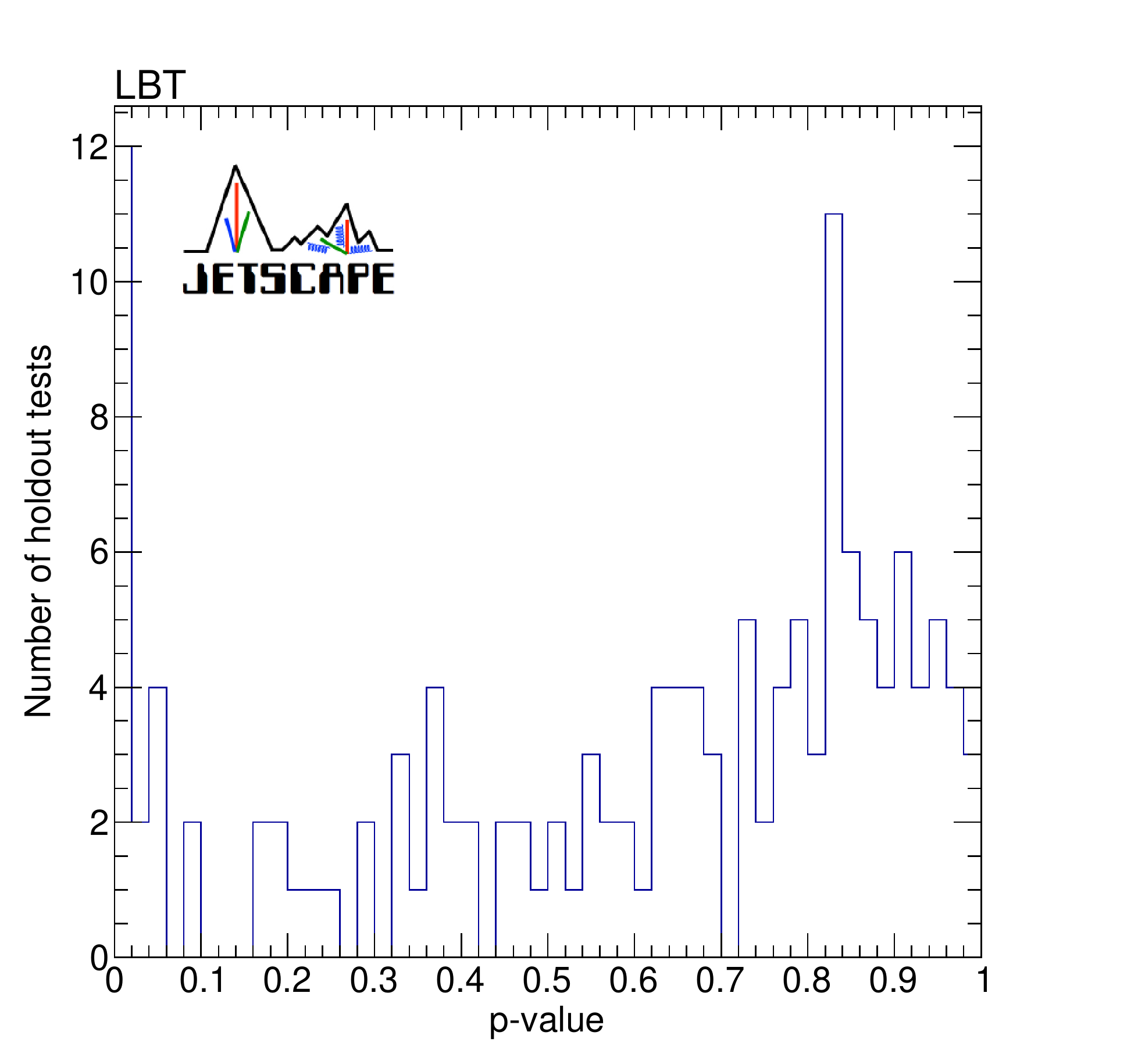}
   \caption{Distribution of $p$-values from the closure tests performed using all the design points.}
   \label{fig:pvalue-lbt}
\end{figure}

Figure \ref{fig:closure-example} shows an example of such a closure test for a single design point, in which the inferred credible region for \qhat\ is compared to 
the true value from the design point. 
We repeat these closure tests for each design point, and statistically evaluate their consistency using a $p$-value.
The $p$-value is defined as the percentage of posterior samples that are more compatible with the pseudo-data than the truth, using a $\chi^2$ taking into account the correlation in the uncertainties.
We generally find consistent performance, as can be seen as an example in Fig.~\ref{fig:pvalue-lbt}.
The distribution deviates from a flat distribution with a shift toward high $p$-values, indicating that the uncertainty obtained is conservative.

Furthermore, we examine the closure differentially in \qhat\ and $\theta$. As described in Sect.~\ref{subsec:emulatorValidation}, we observe consistency except for occasional failure at the boundary of the parameter space,  which are found not to be near the extracted solutions.  The boundary points consist of those with $p$-value very close to 0.
One caveat is that for the \Matter+\Lbt2 multi-stage model, the closure appears to be inconsistent for large values of $\qhat/T^3 \gtrsim 5$. This becomes relevant in the low momentum region, and accordingly the results in that region should be interpreted cautiously. 


\section{Results}
\label{sec:results}

In this section we discuss the posterior distribution from the Bayesian parameter extraction for the parametrizations in Sec.~\ref{sec:qhatParametrization}, and the corresponding values of \qhat. We first discuss the analysis using the \Matter\ and \Lbt\ models separately, and then the analysis of the combined model of \Matter+\Lbt\ with two different choices of \qhat\ parametrization.


\subsection{Parameter extraction using MATTER and LBT separately}
\label{subsec:result-separate}

We first carry out Bayesian parameter extraction for \Matter\ and \Lbt, using  Eq.~(\ref{eq:parametrization1}). 
Figure~\ref{fig:prior-lbt} shows the distribution of inclusive hadron \RAA\ for the three measured datasets, compared to calculations based on \Lbt\ at the initial design points prior to parameter extraction. The prior distribution of the parameter space covers all experimental data and serves as the training data for the GPE. The analogous distributions for the other calculations discussed below look similar to  Fig.~\ref{fig:prior-lbt} and will not be shown.

\begin{figure*}[tbp]
        \centering
                \includegraphics[width=0.9\linewidth]{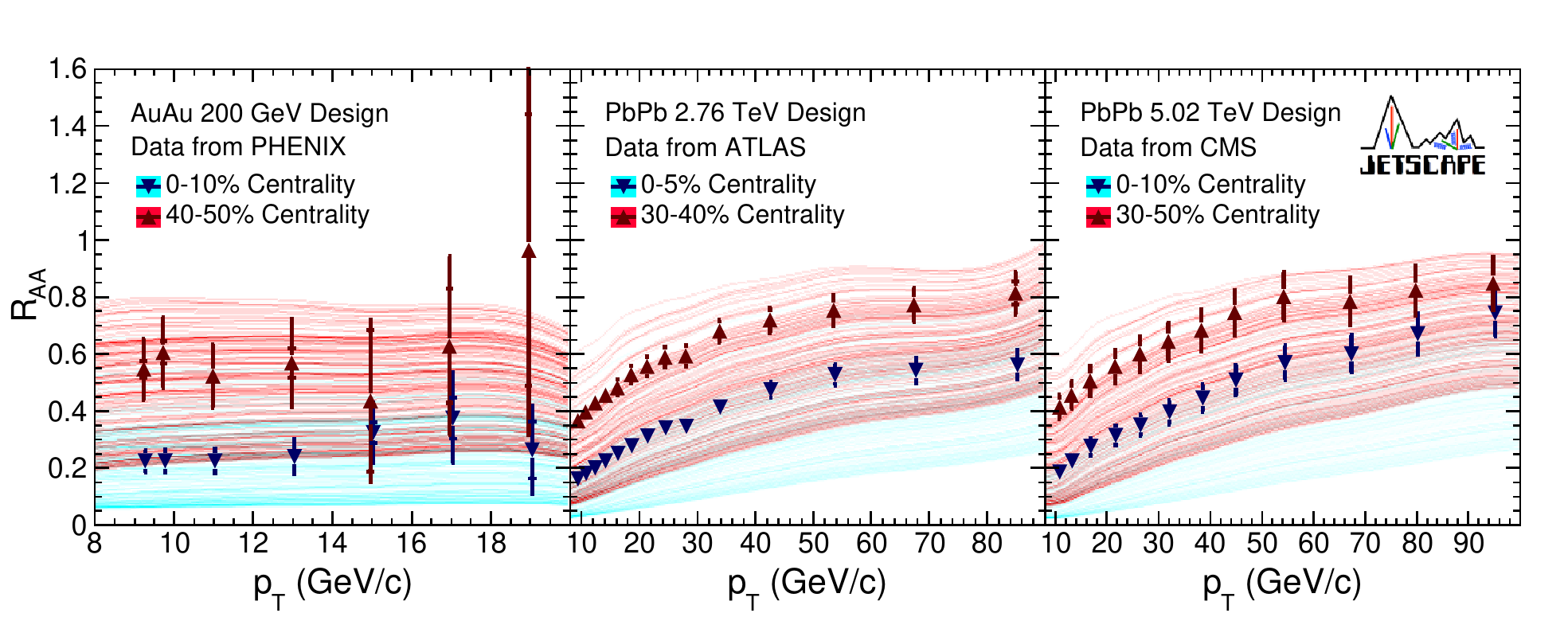}
        \caption{(Color online) Inclusive hadron \RAA\ for the three measured datasets~\cite{Adare:2012wg, Aad:2015wga,Khachatryan:2016odn}, together with prior calculations based on \Lbt\ using design points of the parameter space. Inner error bars on experimental data points are statistical errors; outer error bars are the quadrature sum of statistical error and systematic uncertainty.}
        \label{fig:prior-lbt}
\end{figure*}

\begin{figure*}[tbp]
        \centering
                   \includegraphics[width=0.9\linewidth]{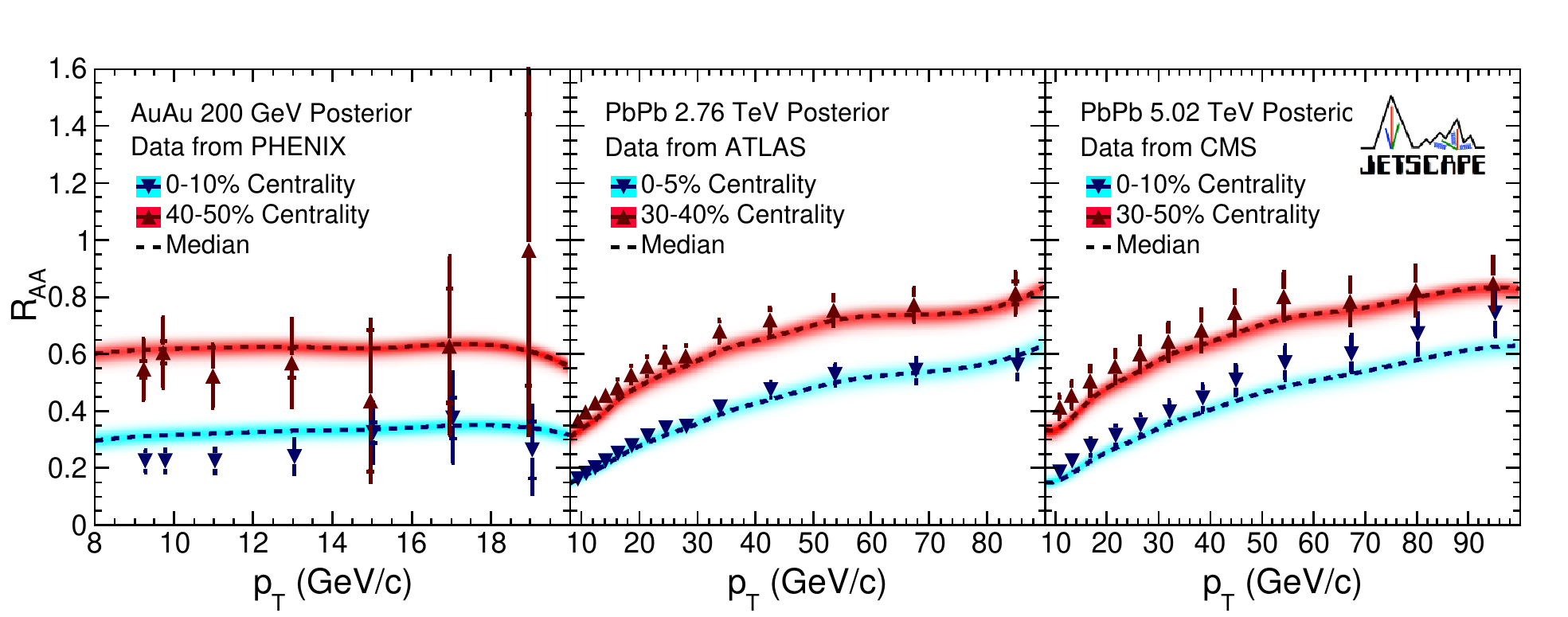}
        \caption{(Color online) Posterior predictive distributions of inclusive hadron \RAA\ using \Lbt\  compared to the same data as Fig.~\ref{fig:prior-lbt}.}
        \label{fig:posterior-lbt}
\end{figure*}

\begin{figure*}[tbp]
        \centering
                \includegraphics[width=0.9\linewidth]{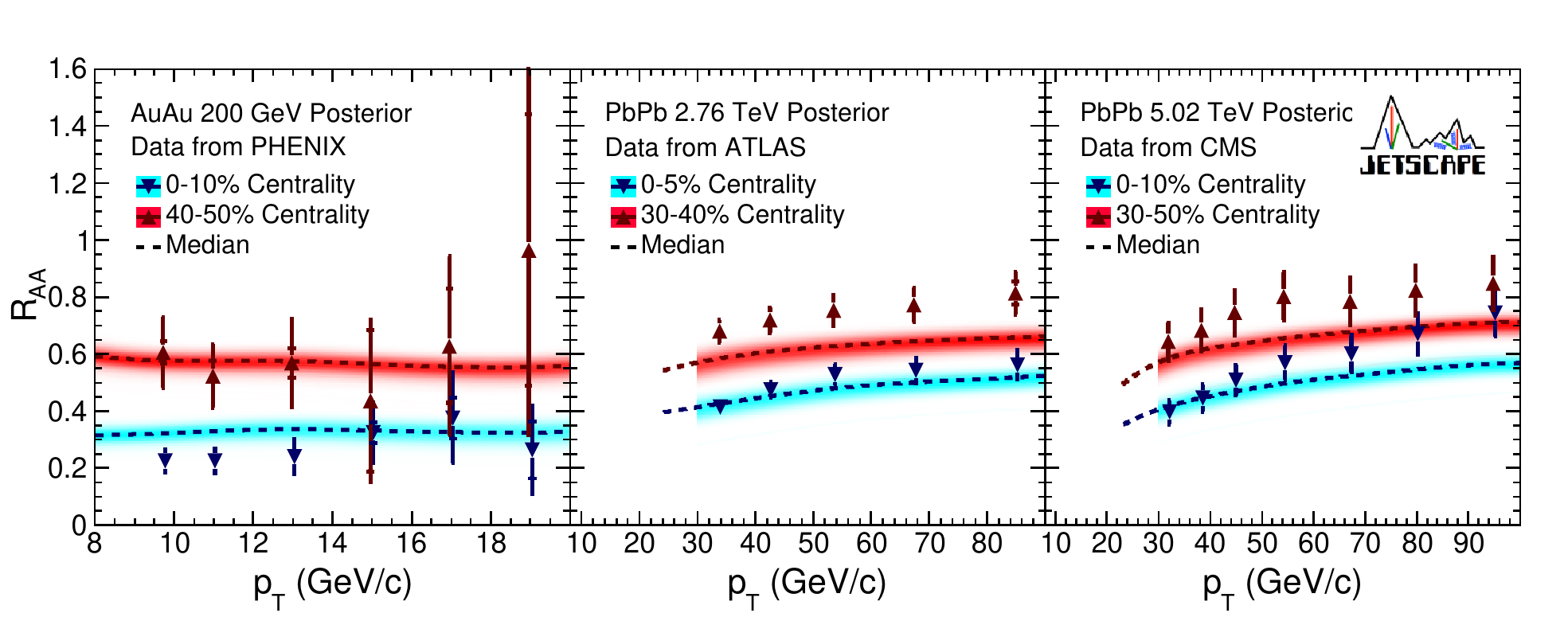}
        \caption{(Color online) Posterior predictive distributions of inclusive hadron \RAA\ using \Matter\ compared to the same data as Fig.~\ref{fig:prior-lbt}.  Data points at lower \pT\ values are excluded from this comparison due to the applicability of the model.}        
        \label{fig:posterior-matter}
\end{figure*}

Figures~\ref{fig:posterior-lbt} and \ref{fig:posterior-matter} show the same data and the posterior distributions for \Lbt\ and \Matter. The dashed lines indicate median values, corresponding to the median parameters values given in Tab.~\ref{tab:median}. The models describe the data moderately well compared to the experimental uncertainties, but exhibit systematic deviations at high \pT\ in central \PbPb\ collisions at \sqrtsNN=5.02 TeV, and at all \pT\  for semi-central collisions in both \sqrtsNN=2.76 and 5.02 TeV.

We compare \Matter\ results with the LHC data (Fig.~\ref{fig:posterior-matter}) only above $p_\mathrm{T}\sim 30$ \gev\ because, when \Matter\ is applied alone, after the first few splittings the parton virtuality may drop below \Qzero\ while its energy is still high. Such partons, when modeled by \Matter, do not interact further with the medium and therefore yield an inaccurate description of the \pT\ dependence of \RAA. \Matter\ is therefore expected to work well at high \pT\ and to fail at low \pT. However, experimental data uncertainties are smaller at low \pT\ than high \pT, and the low \pT\ data therefore contribute with higher weights to the calibration. This generates the deviations between \Matter\ and data at high \pT\ for the LHC data in (Fig.~\ref{fig:posterior-matter}). The implementation in the Bayesian inference analysis of a confidence measure for a model calculation in different regions of phase space will be explored in a future study to address this issue.

\begin{figure}[tb]
        \centering
        \includegraphics[width=0.95\linewidth]{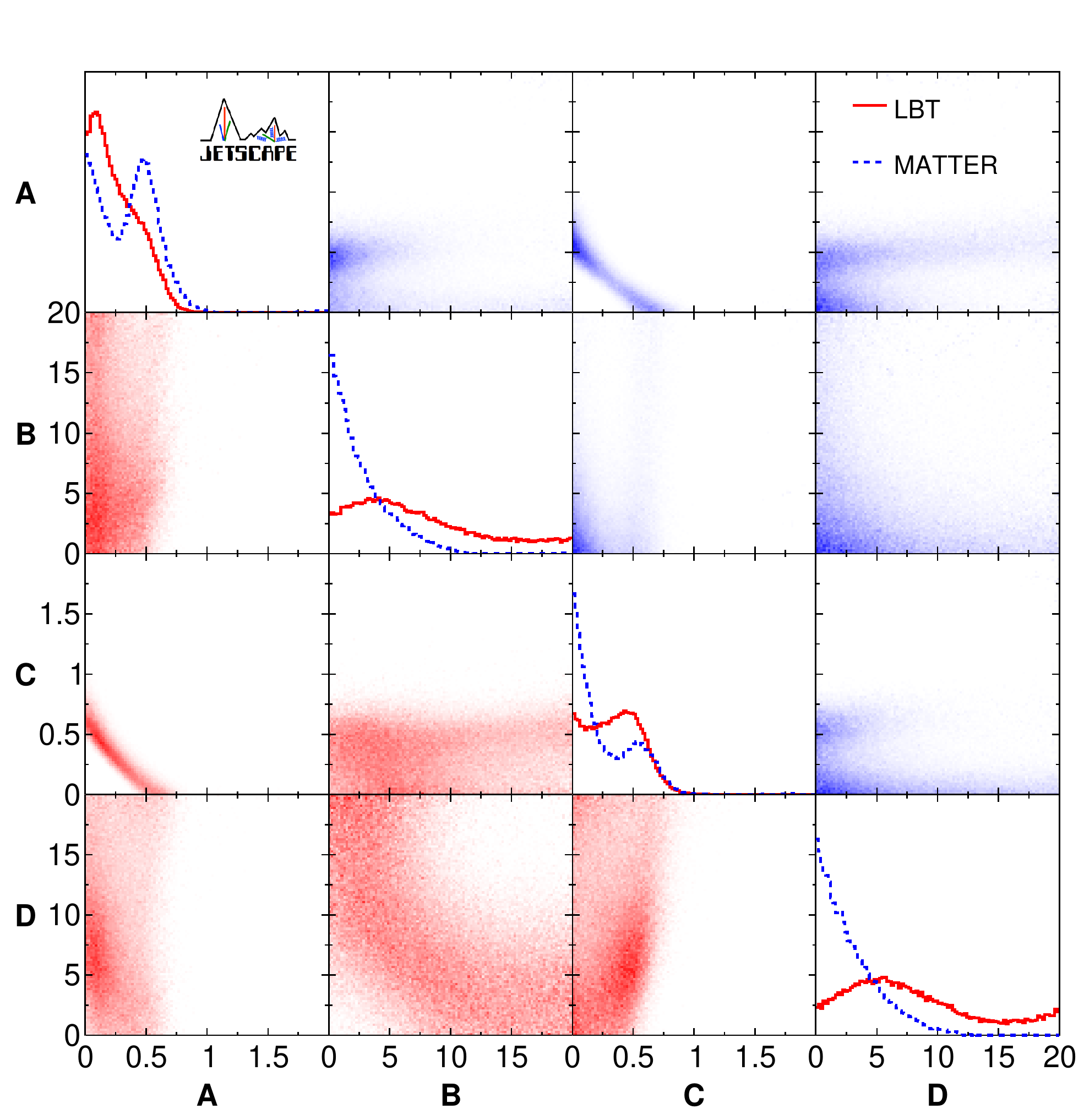}
        \caption{(Color online) Posterior distribution of the 4-D space for \qhat\ when \Matter\ and \Lbt\ are applied separately. Off-diagonal panels show correlations of posterior distributions for \Lbt\ (lower left, red) and  \Matter\ (upper right, blue). To compare the distributions for the two models, parameters A and C are on similar scales, as are B and D. }
        \label{fig:pairplot_separate}
\end{figure}

Figure~\ref{fig:pairplot_separate} shows the posterior distribution of the parameter space from this procedure, with the median value of each parameter given in Tab.~\ref{tab:median}.
The diagonal panels show 1-D projections onto each  parameter;  a clear difference can be seen between the \Matter\ and the \Lbt\ models.
The off-diagonal panels show 2-D projections for \Matter\ (upper right) and \Lbt\ (lower left).

\begin{table}[tb]
\begin{tabular}{ | c | c | c | c | c | c |}
\hline
Parameter & $A$ & $B$ & $C$ & $D$ & $Q$  \\
\hline \hline 
\Matter\ &  0.386 & 3.03 & 0.197 & 3.81 & --   \\ 
\hline
\Lbt\ & 0.225 & 7.20 & 0.354 & 7.95 & -- \\ 
\hline
\Matter+\Lbt~1 & 0.130 & 2.39 & 0.151 & 2.78 & 2.02 \\ 
\hline
\Matter+\Lbt~2 & 0.247 & -- & 0.428 & 6.38 &  2.70 \\ 
\hline
\end{tabular}
\caption{Median values of posterior parameter distributions for the various model parametrizations.
Note that the median values do not take account of correlations between the parameters.}
\label{tab:median}
\end{table}

For \Matter, the extracted value of $A$ is significant while that of $C$ peaks close to zero, indicating
that the extracted value of \qhat\ is due primarily to the first term in the braces in Eq.~(\ref{eq:parametrization1}). In contrast,  \Lbt\ results in a more significant contribution from the second term in Eq.~(\ref{eq:parametrization1}). This is consistent with the respective domains of applicability of the two models: parton splitting inside \Matter\ is driven by high virtuality and is insensitive the thermal scale of the medium, while \Lbt\ describes the scatterings between jet partons with a thermal medium with an on-shell approximation. Note that the domain of experimental data was restricted according to the expected regime of validity of each model.

Figure~\ref{fig:pairplot_separate} also shows the correlation between pairs of parameters in the off-diagonal panels. A marked anti-correlation between parameters $A$ and $C$ is observed in the first column of the third row, because both $A$ and $C$ contribute positively to the overall normalization of \qhat\ in this parametrization. On the other hand, a weaker correlation is seen between $B$ and $D$, which is also expected in  this parametrization.

\begin{figure}[tbp]
        \centering   
        \subfigure{\includegraphics[width=0.8\linewidth]{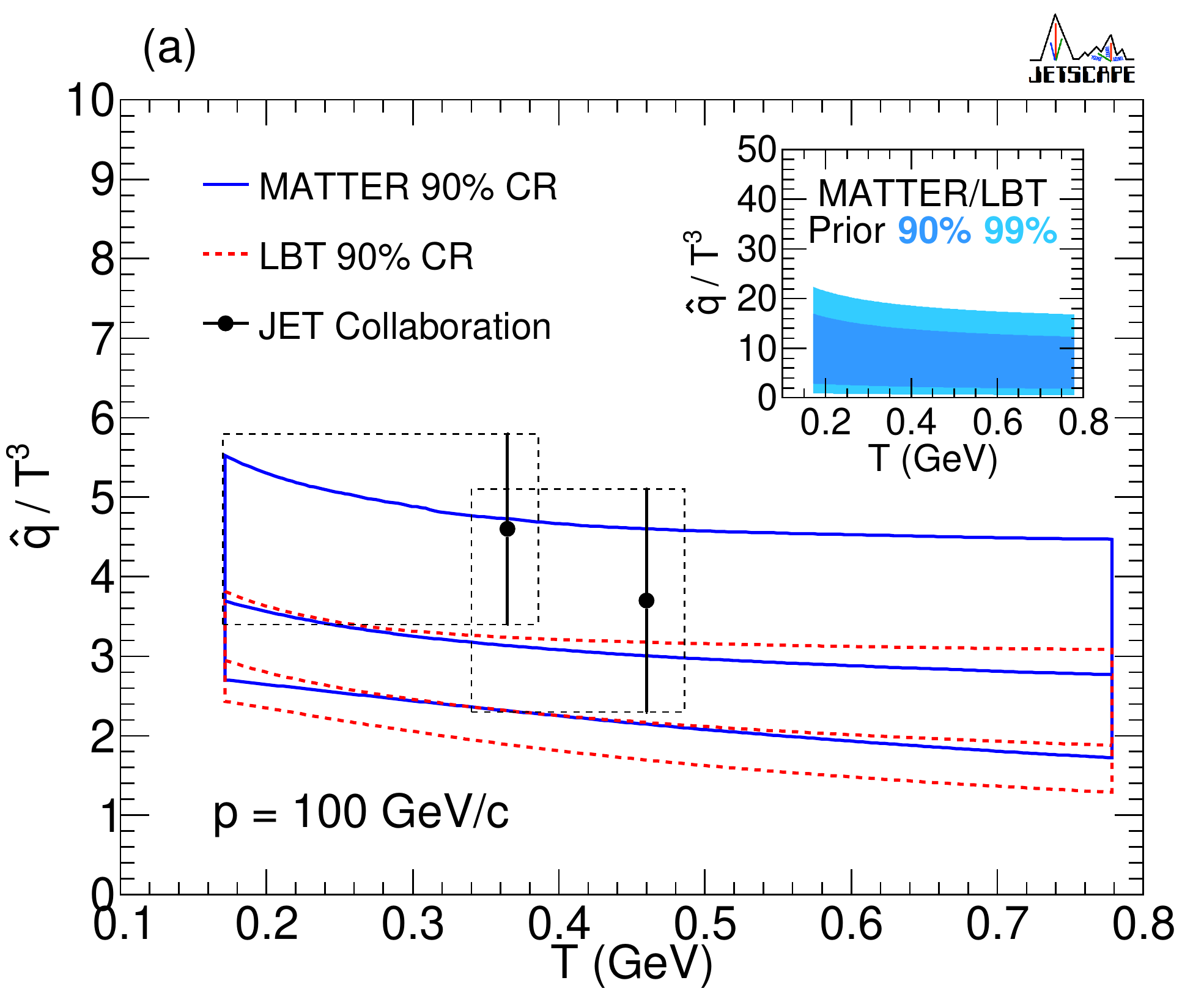}\label{fig:Plot_qhat_T_separate}}
        \subfigure{\includegraphics[width=0.8\linewidth]{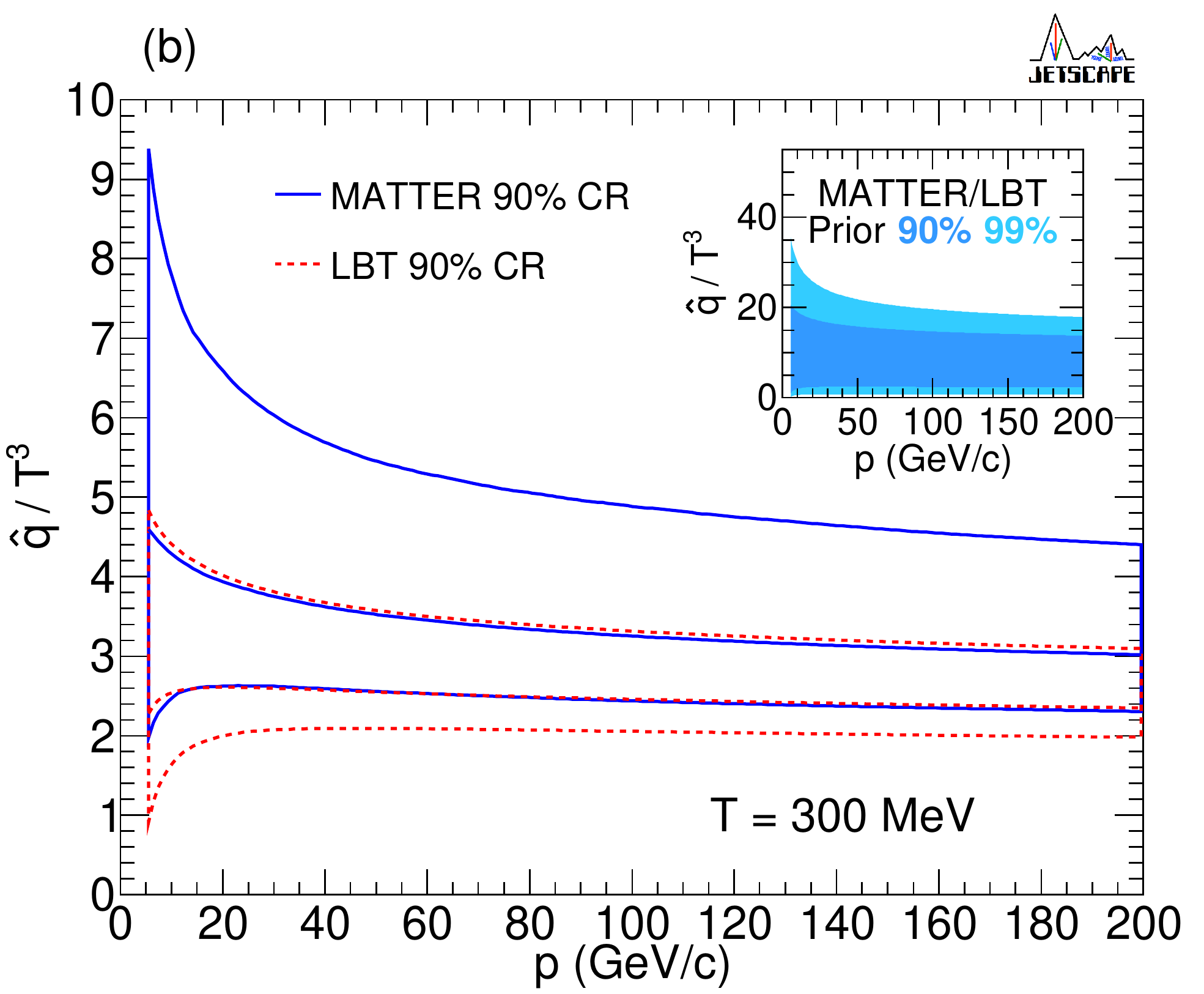}\label{fig:Plot_qhat_p_separate}}                  
        \caption{(Color online) The (quark) jet transport coefficient \qhat\ from Bayesian parameter extraction using \Matter\ and \Lbt\ separately: (a) as function of the medium temperature, and (b) as function of quark momentum. The solid and dashed lines indicate the median value for \Matter\ and  \Lbt, respectively. }
        \label{fig:qhat_separate}
\end{figure}

Figure~\ref{fig:qhat_separate} shows the 90\% credible region (C.R.) for \qhat, determined from the posterior distributions in Fig.~\ref{fig:pairplot_separate}. The dotted and solid lines show the median values for fixed quark momentum and medium temperature in the upper and lower panels, respectively, illustrating more differential information than in Tab.~\ref{tab:median}. This new constraint on \qhat\ is consistent within uncertainties with the value determined  previously by the JET Collaboration~\cite{Burke:2013yra}, although the median value is smaller.  This is expected, since the semi-analytical calculations used in the JET Collaboration analysis did not include elastic scattering processes, and some calculations considered only single gluon emission for the inelastic process. The inclusion of multiple gluon emission channels and elastic scattering in the both \Matter\ and \Lbt\ reduces the extracted \qhat\ value relative to these simpler approximations. 

The extracted value of $\qhat/T^3$ has only weak $T$-dependence for both the \Matter- and \Lbt-based analyses.
Figure~\ref{fig:Plot_qhat_p_separate} shows a slight decrease in $\qhat/T^3$ at high jet \pT\ for both \Matter\ and \Lbt. The uncertainty is larger at low \pT\ due to the \pT-range of \RAA\ data considered in this work.


\subsection{Parameter extraction using MATTER+LBT combined: five parameter}
\label{subsec:result-combine-five}

Figure~\ref{fig:posterior-combine1} shows inclusive hadron \RAA\ with posterior parameter distributions for the combined \Matter+\Lbt~1 model which incorporates the parameterization of \qhat\ in Eq.~\ref{eq:parametrization1} and the switching virtuality \Qzero, for a total of five parameters. The bands show the posterior predictive distributions and the dashed lines are results from using the median parameter values listed in Tab.~\ref{tab:median}. Compared to the fits with \Matter\ or \Lbt\ alone, there is no significant improvement when fitting with the combined model. 
The level of agreement of the posterior distributions is similar to that seen in Figs.~\ref{fig:posterior-lbt} and \ref{fig:posterior-matter}, indicating that the simple model of a virtuality scale \Qzero\ for switching between \Matter\ and \Lbt\ may not fully capture the virtuality dependence of jet quenching.

\begin{figure*}[tbp]
        \centering
                \includegraphics[width=0.9\linewidth]{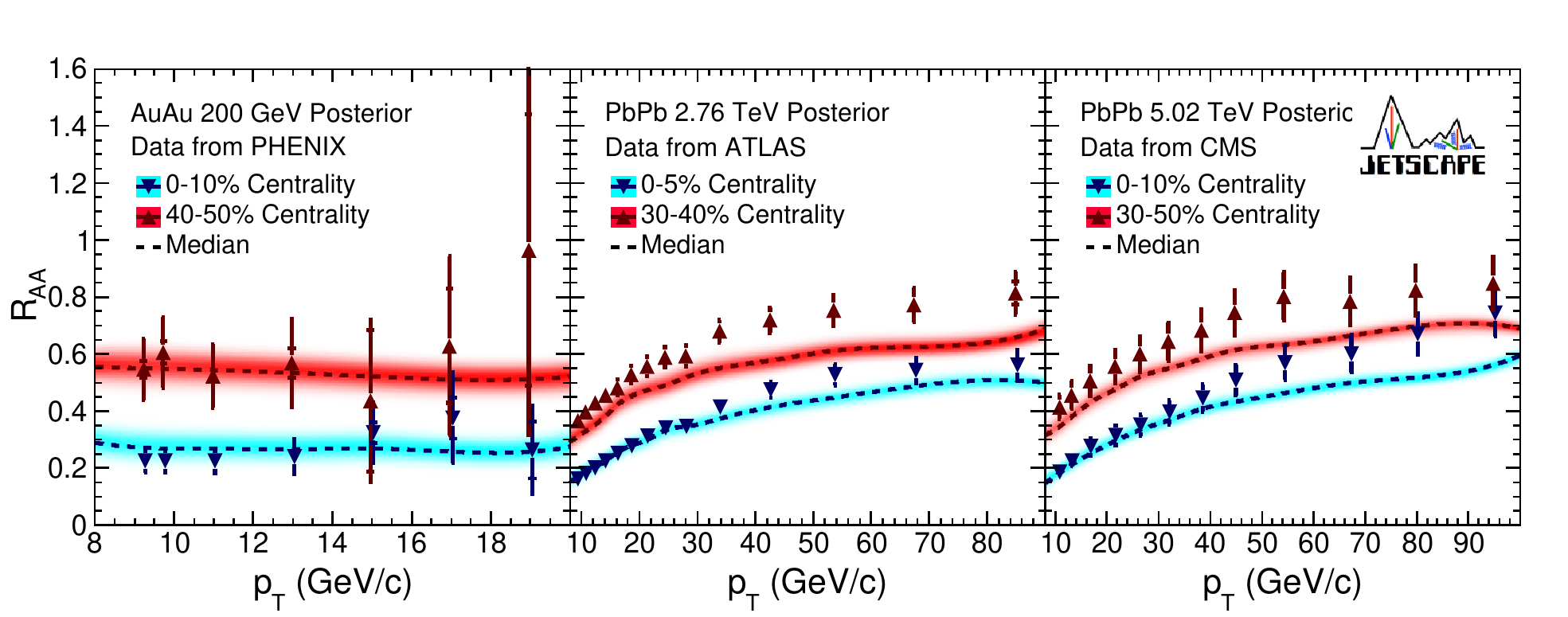}
        \caption{(Color online) Posterior predictive distributions of \Matter+\Lbt~1 compared to the same data as Fig.~\ref{fig:prior-lbt}. }
        \label{fig:posterior-combine1}
\end{figure*}

Figure~\ref{fig:pairplot_combine1} shows the correlation of posterior parameter distributions for the \Matter+\Lbt~1 parametrization. Results are shown separately for fits to the RHIC and LHC data, as well as the combined parameter extraction to all six data sets. Figure~\ref{fig:pairplot_combine1} shows the constraint on \Qzero, the virtuality scale at which the calculation switches between \Matter\ and \Lbt. The median value is 2.02~GeV, with 90\% credible region $[1.25,2.72]$~GeV. 
It is evident that the RHIC data have significantly less impact than the LHC data on the posterior distributions in this analysis, because of the low relative statistical weight of the selected RHIC data (Sect.~\ref{sec:ExpData}).
Line 3 of Table~\ref{tab:median} gives the median values of the five parameters.

\begin{figure}[tbp]
\centering
\includegraphics[width=0.95\linewidth]{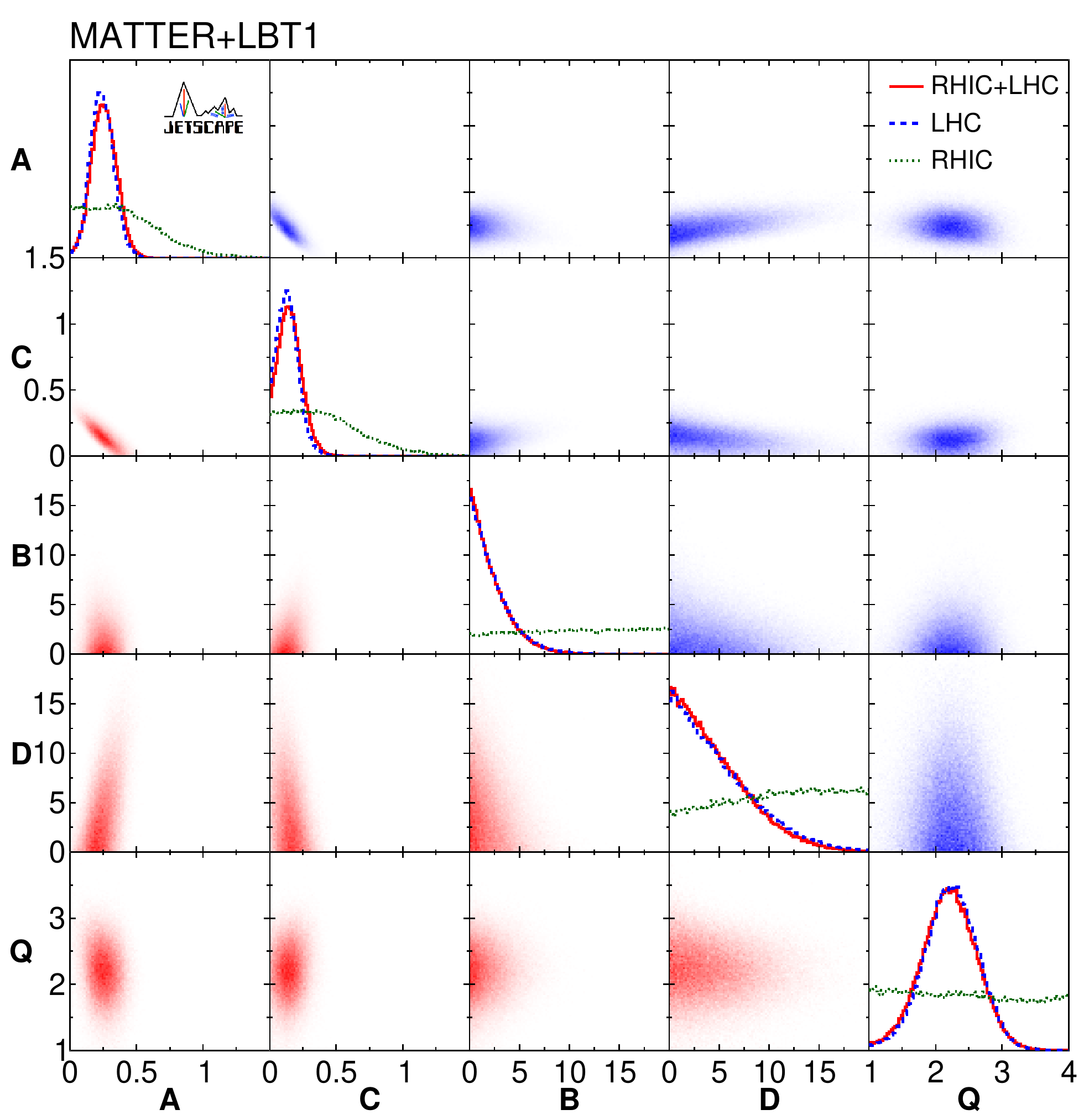}
\caption{(Color online) Posterior distribution of the 4-D space for \Matter+\Lbt 1. Off-diagonal panels show correlations of posterior distributions for RHIC+LHC (lower left, red) and LHC only (upper right, blue). 
}
\label{fig:pairplot_combine1}
\end{figure}

The value of \Qzero\ in this analysis is taken to be the same for RHIC and LHC data, though in practice the value of \Qzero\ may be  smaller at RHIC than at the LHC because of the lower average QGP temperature. Exploration of this degree of freedom requires consideration of additional experimental data, however, and is beyond the scope of the current analysis.

\begin{figure}[tbp]
        \centering
        \subfigure{\includegraphics[width=0.8\linewidth]{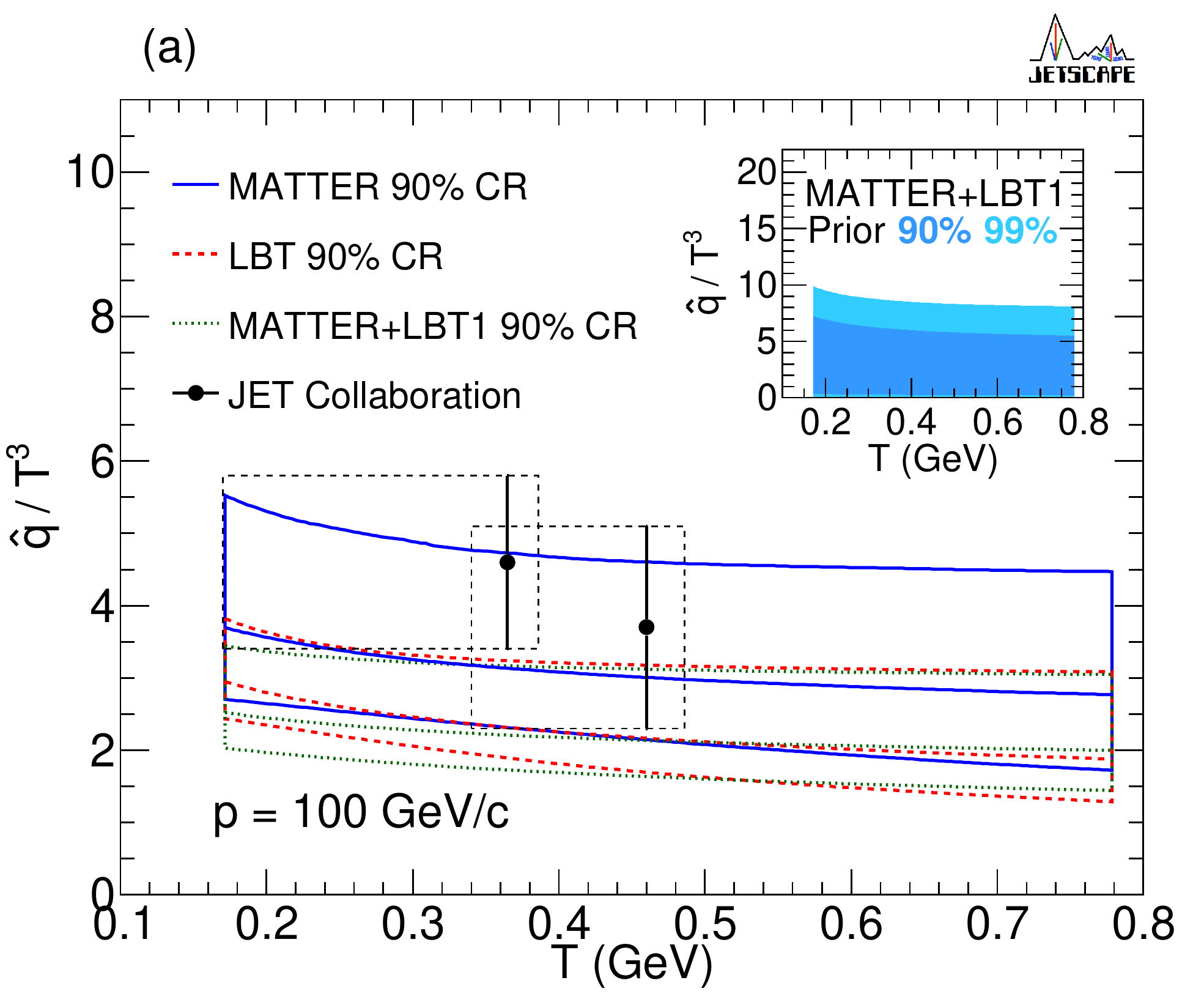}\label{fig:Plot_qhat_T_combine1}}
        \subfigure{\includegraphics[width=0.8\linewidth]{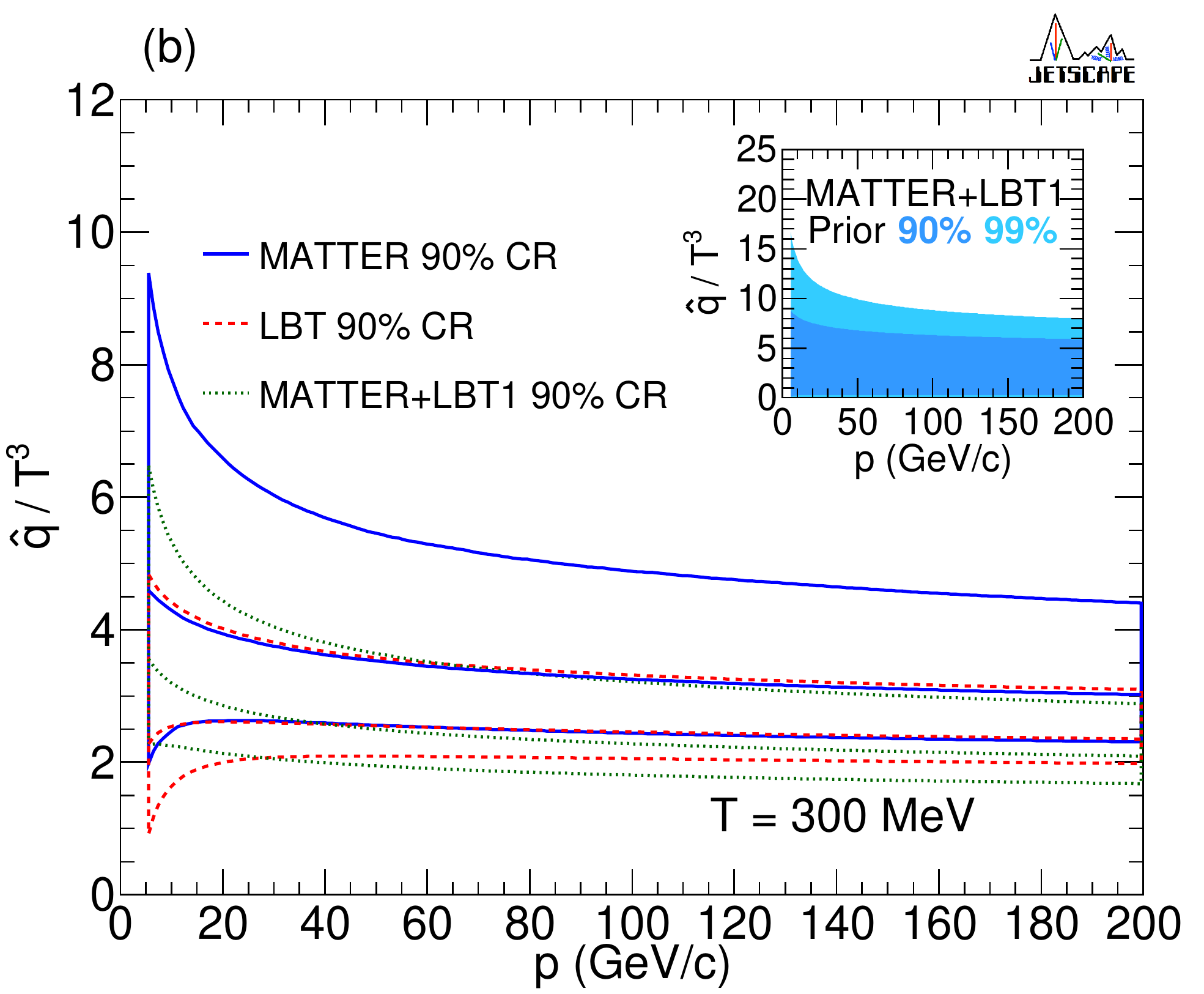}\label{fig:Plot_qhat_p_combine1}}               
        \caption{(Color online) 90\% CR regions for the quark-jet transport coefficient \qhat\ using \Matter\ and \Lbt\ (Fig.~\ref{fig:qhat_separate})  and \Matter+\Lbt 1 (a) as function of medium temperature, and (b) as function of quark energy. The lines at the center of the bands indicate their median values.  The data points (black circles with vertical error bars) show the result from the JET Collaboration~\cite{Burke:2013yra}; dotted boxes indicate the range of that analysis.}
        \label{fig:qhat_combine1}
\end{figure}

Figure~\ref{fig:qhat_combine1} shows the 90\% C.R. of the extracted quark jet \qhat\ from the \Matter+\Lbt~1 parametrization, together with its median value (dashed line). Also shown are the bands for extraction from either \Matter\ and \Lbt\ taken from Fig.~\ref{fig:qhat_separate}. The value of \qhat\ determined using the multi-stage approach is lower than those determined from fits to \Matter\ or \Lbt\ separately. This is because \Matter\ is effective for jet energy loss at high virtuality but is blind to parton evolution at low virtuality, while the opposite is the case for  \Lbt. Combining the models for a multi-stage evolution approach leads to larger jet energy loss than found by applying only one of them across the entire phase space. \Matter+\Lbt\ therefore requires a smaller \qhat\ value than \Matter\ or \Lbt\ does to describe the same jet quenching data.


\subsection{Parameter extraction using MATTER+LBT combined: four parameter}
\label{subsec:result-combine-four}

\begin{figure*}[tbp]
        \centering
                \includegraphics[width=0.9\linewidth]{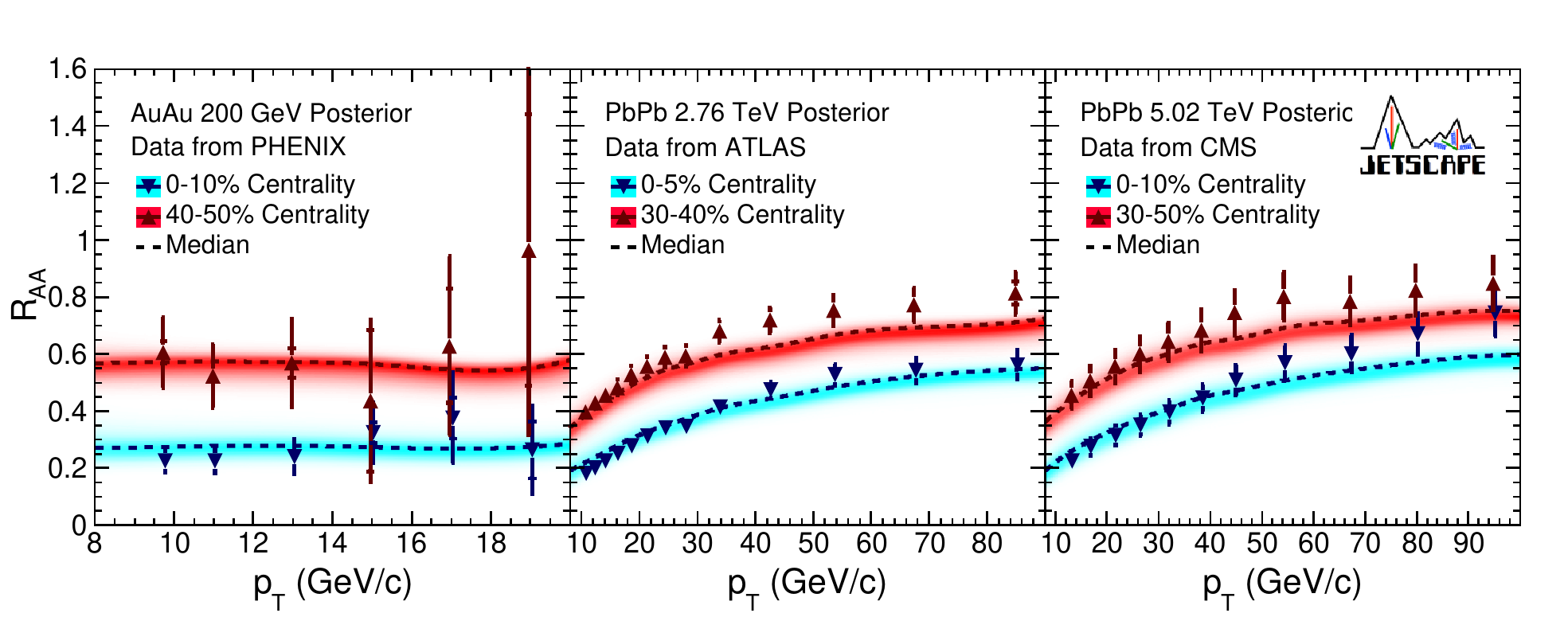}
        \caption{(Color online) Posterior predictive distributions of \RAA\ using \Matter+\Lbt~2, compared to the same data as Fig.~\ref{fig:prior-lbt}. Dashed lines show model calculation using median values of parameters.}
        \label{fig:posterior-combine2}
\end{figure*}

Finally, we discuss results using the \Matter+\Lbt~2 parametrization given in Eq.~\ref{eq:parametrization2}. Figure~\ref{fig:posterior-combine2} shows the posterior \RAA\ distribution and the model calculation utilizing the median values of the parameters. Qualitatively, the model calculations describe the overall \pT-dependence of the data well although, as also seen in Figs.~\ref{fig:posterior-lbt}, \ref{fig:posterior-matter} and \ref{fig:posterior-combine1}, the posterior distributions fall outside of the systematic uncertainty limits of the data. This again indicates that introduction of a virtuality scale \Qzero\ for switching between \Matter\ and \Lbt\ may not be sufficient to describe the virtuality dependence of jet quenching.

\begin{figure}[tbp]
        \centering
               \includegraphics[width=0.95\linewidth]{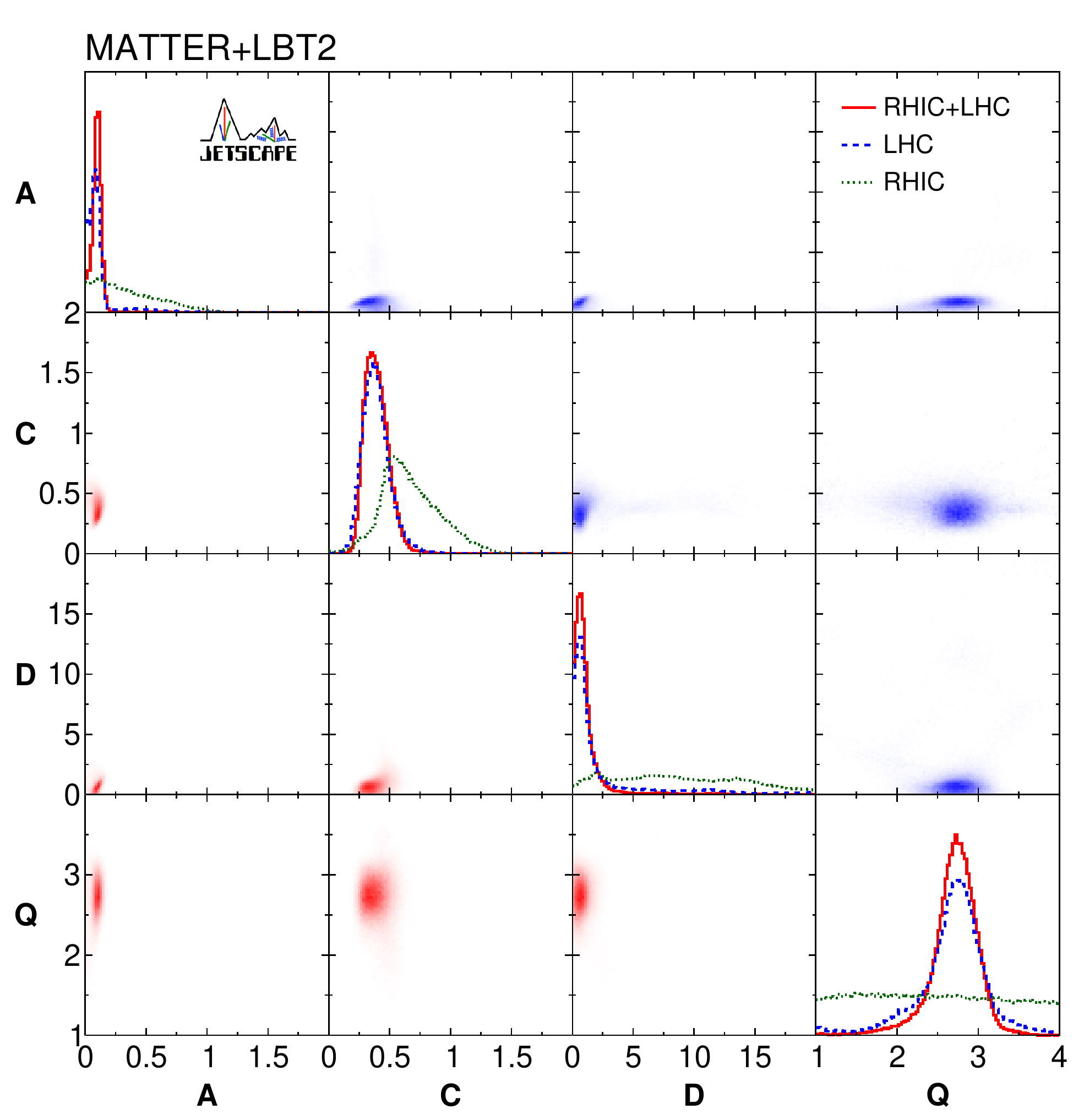}
        \caption{(Color online) Posterior distribution of the 4-D space for \Matter+\Lbt 2. Off-diagonal panels show correlations of posterior distributions for RHIC+LHC (lower left, red) and LHC only (upper right, blue). }
        \label{fig:pairplot_combine2}
\end{figure}

Figure~\ref{fig:pairplot_combine2} shows the correlation of posterior parameter distributions of  \Matter+\Lbt~2 using RHIC and LHC data separately and combined. The posterior parameter distributions are significantly less constrained by the RHIC than LHC data, similar to the case of  \Matter+\Lbt~1 (Fig.~\ref{fig:pairplot_combine1}). The median value of \Qzero\  from combined RHIC and LHC data is 2.70~GeV, with 90\% CR $[1.84,3.41]$~GeV. While the median value is larger than that determined using \Matter+\Lbt~1, the difference is not significant, as seen from the mutually compatible 90\% confidence regions. This comparison provides an estimate of the systematic uncertainty due to different model parametrization. Line 4 of Table~\ref{tab:median} gives the median values of the four parameters.

\begin{figure}[tbp]
        \centering
         \subfigure{\includegraphics[width=0.8\linewidth]{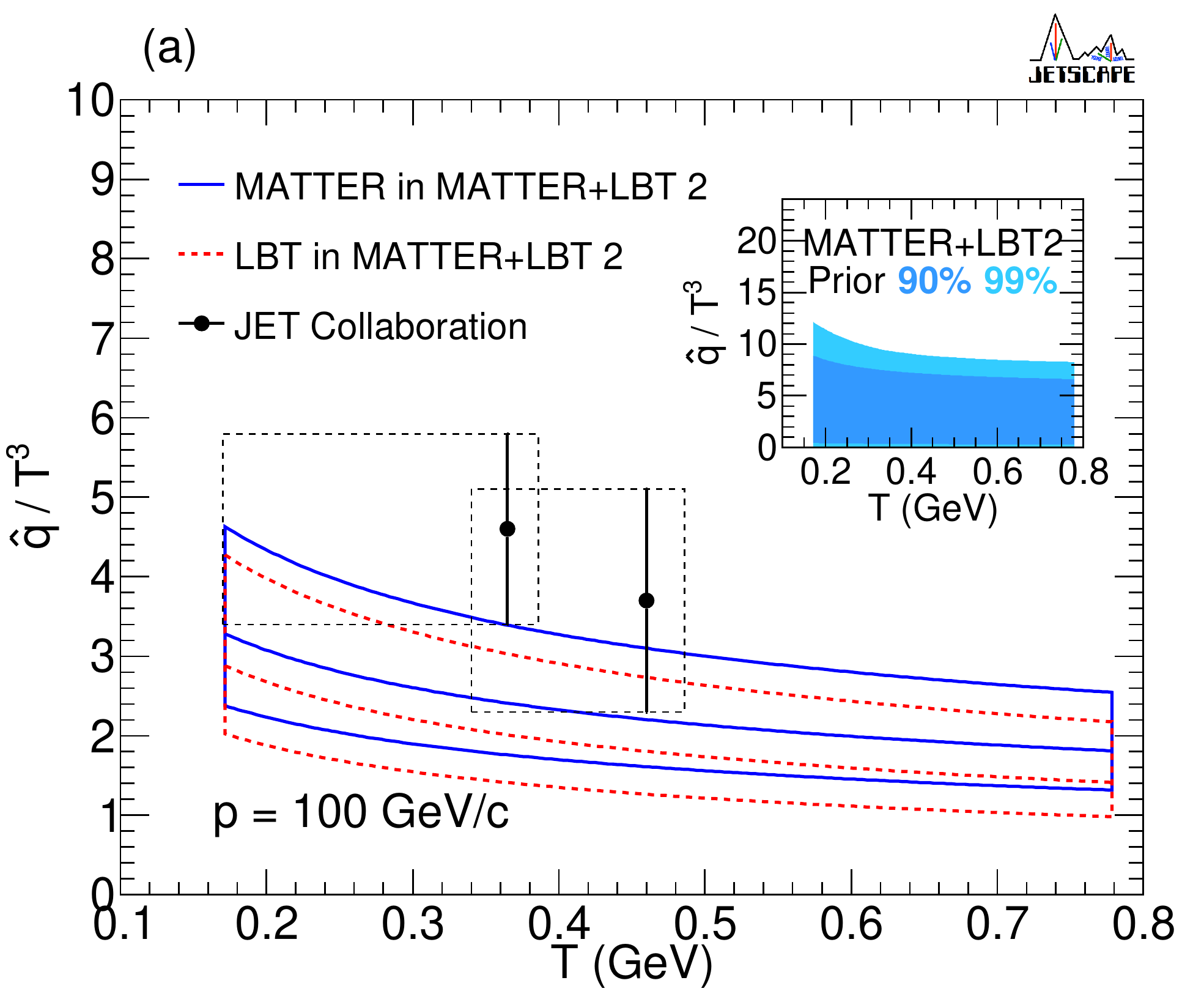}\label{fig:Plot_qhat_T_combine2}}
         \subfigure{\includegraphics[width=0.8\linewidth]{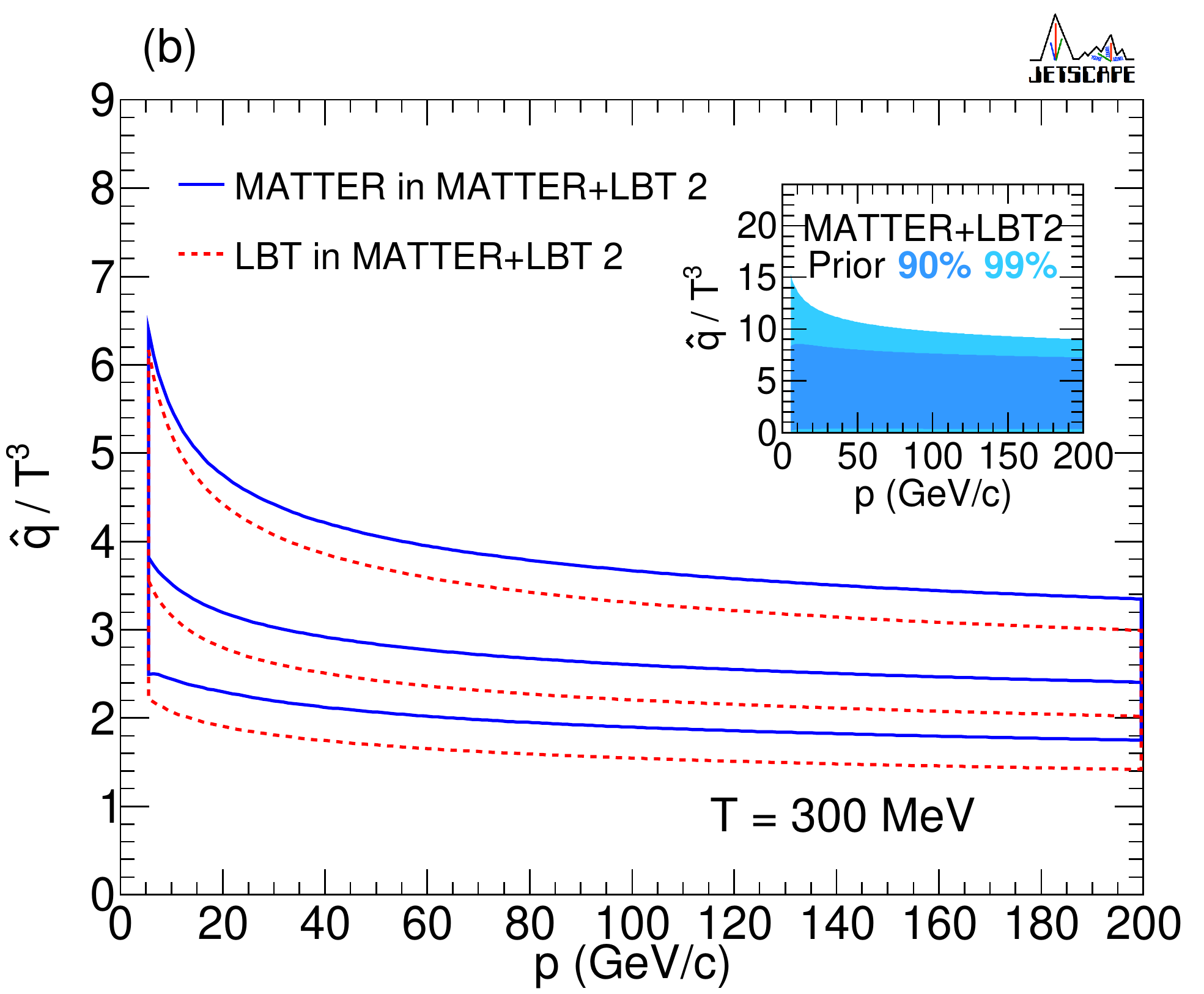}\label{fig:Plot_qhat_p_combine2}}                  
        \caption{(Color online) 90\% CR regions for the quark-jet transport coefficient \qhat\ using \Matter+\Lbt 2 (a) as function of medium temperature, and (b) as function of quark energy. The lines at the center of the bands indicate their median values.  The data points (black circles with vertical error bars) show the result from the JET Collaboration~\cite{Burke:2013yra}; dotted boxes indicate the range of that analysis.}
        \label{fig:qhat_combine2}
\end{figure}

Figure~\ref{fig:qhat_combine2} shows \qhat\ as a function of medium temperature and quark momentum using \Matter+\Lbt~2. Referring to Eq.~(\ref{eq:parametrization2}), \qhat\ has a larger value in the \Matter\ stage than in the \Lbt\ stage, since both terms in $\{\ldots\}$ contribute to \Matter\ but only the second term contributes to \Lbt. The contribution from the first term depends on the virtuality of the parton $Q$, which varies for different partons with the same momentum. 
In order to estimate the \qhat\ value in Fig.~\ref{fig:qhat_combine2} we use the average value of $Q$, obtained from Eq.~(\ref{eq:matter}). Comparing with Fig.~\ref{fig:qhat_combine1}, we also observe that the value of \qhat\ extracted from \Matter+\Lbt~2 is larger than that from \Matter+\Lbt~1.


\section{Summary and Outlook}
\label{sec:summary}

We have reported the application of state-of-the-art Bayesian inference methodology to determine the QGP jet transport coefficient \qhat\ from inclusive hadron suppression data measured at RHIC and the LHC. Two jet energy loss models were utilized, \Matter\ and \Lbt. \Matter\ is applicable to modeling the medium-modified splitting of highly virtual partons, while \Lbt\ is applicable for the in-medium transport of nearly on-shell partons. The models are first applied separately, and then combined to form a multi-stage evolution approach. Two different parametrizations were used for the functional dependence of \qhat\ on jet momentum or virtuality scale and the medium temperature, based on the picture of perturbative scattering between jets and a thermal medium. A novel treatment of experimental uncertainties is employed, taking into account their covariance for the first time in the determination of \qhat.

Such model calculations are computationally expensive. Gaussian process emulators are therefore employed to render this process computationally efficient, trained at design points selected starting with a Latin Hypercube in  parameter space. The resulting gain in computational efficiency enabled calibration of the multi-dimensional parameter space.

The Bayesian inference process generates posterior parameter distributions for each model configuration. 
To constrain the model parameters, we used 66 inclusive hadron \RAA\ datapoints at two centralities for \AuAu\ collisions at \sqrtsNN=200~GeV and \PbPb\ collisions at \sqrtsNN=2.76 and 5.02 TeV. 
For both the \Matter-only and \Lbt-only configurations, the extracted value of $\hat{q}/T^3$ has only weak dependence on the medium temperature $T$.
The value of \qhat\ determined using these approaches is consistent with a previous determination by the JET Collaboration~\cite{Burke:2013yra}. 

A multi-stage jet evolution approach, combining \Matter\ and \Lbt, is applied here for the first time. The transition between \Matter\ and \Lbt, based on parton virtuality, is controlled by the virtuality parameter \Qzero\ which separates the virtuality-ordered-splitting dominating region and the time-ordered-transport dominating region for jet quenching inside a medium. The posterior distribution of $\qhat/T^3$ from the combined model (\Matter+\Lbt) is systematically lower than that determined using \Matter\ or \Lbt\ alone, since the combined model more accurately describes energy loss over the full virtuality range, with similar \pT-dependence. The  two different \qhat\ parametrizations give consistent results, although with differences in the median extracted parameter values; the median value of \Qzero\ is 2.0 and 2.7\,GeV, respectively.

The application of Bayesian inference in this analysis represents a significant advance in quantitative understanding of jet-medium interactions in the Quark-Gluon Plasma. However, the posterior distributions from this analysis do not fully describe the magnitude and \pT-dependence of inclusive hadron \RAA\ measurements in the datasets considered. This tension indicates that additional components in the modeling of jet quenching are needed, for instance a more detailed parametrization of the virtuality dependence of jet quenching than the single switching scale \Qzero\ used here.

Future work will also provide more detailed accounting of experimental and theoretical uncertainties and their covariance and incorporate additional measurement channels, in particular those involving coincidence observables and reconstructed jets. While the  \qhat\ parametrization employed in this analysis is derived from the perturbative approach to jet-medium scattering, non-perturbative effects may require additional dependence of \qhat\ on jet energy and medium temperature, and additional transport parameters may be required to fully describe the jet measurements. 
In addition, aspects of modeling the hydrodynamic medium that have not been considered in this analysis will be explored. Other models of the plasma, incorporating quasi-particle degrees of freedom~\cite{Das:2015ana,Xu:2014tda}, will also be considered.


\section*{Acknowledgments} 
This work was supported in part by the National Science Foundation (NSF) within the framework of the JETSCAPE collaboration, under grant numbers ACI-1550172 (Y.C. and G.R.), ACI-1550221 (R.J.F., F.G., M.K. and B.K.), ACI-1550223 (D.E., M.M., U.H., and L.D.), ACI-1550225 (S.A.B., J.C., T.D., W.F., R.W., S.M., and Y.X.), ACI-1550228 (J.M., B.J., P.J., W.K., X.-N.W.), and ACI-1550300 (S.C., L.C., A.K., A.M., C.N., C.P., A.S., J.P., L.S., C.Si., R.A.S. and G.V.). It was also supported in part by the NSF under grant numbers OAC--2039142 (R.A.), PHY-1516590, PHY-1812431 and PHY-2012922 (R.J.F., B.K., F.G., M.K., and C.S.), and by the US Department of Energy, Office of Science, Office of Nuclear Physics under grant numbers \rm{DE-AC02-05CH11231} (D.O., X.-N.W.), \rm{DE-AC52-07NA27344} (A.A., R.A.S.), \rm{DE-SC0013460} (S.C., A.K., A.M., C.S. and C.Si.), \rm{DE-SC0004286} (L.D., M.M., D.E. and U.H.), \rm{DE-SC0012704} (B.S. and C.S.), \rm{DE-FG02-92ER40713} (J.P.) and \rm{DE-FG02-05ER41367} (T.D., J.-F.P., S.A.B. and Y.X.). The work was also supported in part by the National Science Foundation of China (NSFC) under grant numbers 11935007, 11861131009 and 11890714 (Y.H. and X.-N.W.), by the Natural Sciences and Engineering Research Council of Canada (C.G., M.H., S.J., C.P. and G.V.), by the Fonds de recherche du Qu\'{e}bec -- Nature et technologies (FRQNT) (G.V.), by the Office of the Vice President for Research (OVPR) at Wayne State University (C.P. and Y.T.), by the S\~{a}o Paulo Research Foundation (FAPESP) under projects 2016/24029-6, 2017/05685-2 and 2018/24720-6 (M.L.), and by the University of California, Berkeley - Central China Normal University Collaboration Grant (W.K.). U.H. would like to acknowledge support by the Alexander von Humboldt Foundation through a Humboldt Research Award. Allocation of supercomputing resources (Project: PHY180035) were obtained in part through the Extreme Science and Engineering Discovery Environment (XSEDE), which is supported by National Science Foundation grant number ACI-1548562. Calculation were performed in part on Stampede2 compute nodes, generously funded by the National Science Foundation (NSF) through award ACI-1134872, within the Texas Advanced Computing Center (TACC) at the University of Texas at Austin \cite{TACC}, and in part on the Ohio Supercomputer \cite{OhioSupercomputerCenter1987} (Project PAS0254). Computations were also carried out on the Wayne State Grid funded by the Wayne State OVPR, and on the supercomputer \emph{Guillimin} from McGill University, managed by Calcul Qu\'{e}bec and Compute Canada. The operation of the supercomputer \emph{Guillimin} is funded by the Canada Foundation for Innovation (CFI), NanoQu\'{e}bec, R\'{e}seau de M\'{e}dicine G\'{e}n\'{e}tique Appliqu\'{e}e~(RMGA) and FRQ-NT.  Data storage was provided in part by the OSIRIS project supported by the National Science Foundation under grant number OAC-1541335.




\bibliographystyle{apsrev4-1}

%

\end{document}